\magnification=\magstep1
\hsize=6truein
\vsize=8.9truein
\baselineskip=16truept

\noindent
\centerline{{\bf Universality of phase transitions of frustrated 
antiferromagnets}}

\bigskip
\noindent
\centerline{Hikaru Kawamura}

\bigskip
\noindent
\centerline{{\sl Faculty of Engineering and Design
}}
\centerline{{\sl Kyoto Institute of Technology,
Sakyo-ku, Kyoto 606, Japan}}
\par
\bigskip
{\narrower
Recent theoretical and experimental studies on the
critical properties of
frustrated antiferromagnets with the noncollinear spin order,
including stacked-triangular antiferromagnets and helimagnets,
are reviewed. Particular emphasis is put on the 
novel critical and multicritical 
behaviors exhibited by these magnets,
together with
an important role played by the `chirality'.
\par}
\bigskip
\medskip
\noindent
{\bf \S 1. Introduction}
\medskip
Phase transitions and critical phenomena have been a
central issue of statistical physics for many years.
In particular, phase transitions of magnets or of
`spin systems' have attracted special interest.
Thanks to extensive theoretical and experimental studies,
we now have rather good understanding of the nature of phase
transitions of standard ferromagnets and antiferromagnets.
By the term `standard', I mean here regular and unfrustrated
magnets without quenched disorder and frustration. They include
ferromagnets 
and
unfrustrated antiferromagnets with the collinear spin
order.
\par
One key notion
which  emerged through these studies is the notion of 
universality. 
According to the universality hypothesis, 
a variety of continuous (or second-order) 
phase transitions can be  classified
into a small number of universality classes determined by a few basic
properties characterizing the system under study, 
such as the space dimensionality $d$,
the symmetry of the order parameter and the range of interaction.
If one is interested only in the so-called 
universal quantities, such as
critical exponents, amplitude ratios and scaled equation
of state, various phase transitions should exhibit 
exactly the same behavior. In the case of standard bulk magnets
in three spatial dimensions ($d=3$), universality class is basically 
determined by the number of the spin components, $n$.
Physically, the index $n$ 
is related to the type of magnetic anisotropy:
Namely, $n=1$ (Ising), $n=2$ ({\it XY\/}) and
$n=3$ (Heisenberg) correspond to magnets with
easy-axis-type anisotropy, easy-plane-type anisotropy and no
anisotropy (isotropic magnets), respectively.
The critical properties associated with these $n$-component 
$O(n)$ universality classes 
have been extensively studied and are now rather well understood.
From the renormalization-group (RG) 
viewpoint, these critical 
properties are governed by the so-called Wilson-Fisher 
$O(n)$ fixed point.

Of course, there are a class of magnets exhibiting phase 
transitions very different from the standard $O(n)$ behavior.
One such example may be seen in random magnets 
with quenched disorder. A typical example of such random magnets is 
a spin glass, a magnet not only random but also {\it frustrated\/}.
Even in regular magnets without quenched disorder,
one could expect novel transition 
behavior if the magnets are frustrated.
In fact, the nature of phase transitions
of frustrated magnets could be  novel and
entirely different from those of conventional 
unfrustrated magnets as we shall see in what follows. 
\par\bigskip
\noindent
(a) Frustration
\par\medskip
Frustration could arise either from the 
special geometry of the lattice,
or from the competition between the near-neighbor and further neighbor
interactions. 
The former type of frustration may be seen in
antiferromagnets on a two-dimensional (2D) triangular lattice or
on a 
three-dimensional (3D) stacked-triangular (simple hexagonal) lattice,
which consists of 
two-dimensional triangular layers 
stacked along an orthogonal direction.
The latter type of frustration may be realized in 
helimagnets where magnetic
spiral is formed along a certain direction of the lattice.
\par
Spin frustration brings about interesting
consequences on the resulting
spin structures. As an example, let us consider
three antiferromagnetically-coupled spins located
at each corner of a triangle.
The stable spin configurations differ 
depending on the type of spin symmetry, or the number of spin 
components $n$.
In the case of one-component Ising spins ($n=1$), 
the ground state is not uniquely determined: The situation here is 
illustrated in Fig.1.
Frustration in the Ising case thus leads to
the nontrivial
degeneracy of the ordered state. 

By contrast, when the spin has a continuous symmetry
as in the case of vector spins such as
the two-component {\it XY\/} ($n=2$) and 
the three-component Heisenberg ($n=3$) spins, the ground-state spin
configurations become
{\it noncollinear\/} or {\it canted\/}, as illustrated in Fig.2.
Note that, 
in this case, frustration is partially released by
mutual spin canting and there no longer remains a nontrivial
degeneracy of the ground state up to  global $O(n)$ spin rotation
and reflection.
In this article, 
we shall concentrate on this
latter type of frustrated magnets with the noncollinear or canted
ordered states.
\par\bigskip
\noindent
(b) Chirality
\par\medskip
One interesting consequence of such canted spin structures
is the appearance of a `chiral' degree of freedom.
Let us consider, for example,  the case of  {\it XY\/} spins
shown in Fig.2. 
If the  exchange interactions are 
equal in magnitude on the three bonds,
the ground-state spin configuration 
is the so-called 
`120$^\circ $ spin 
structure', in which three {\it XY\/} spins 
form 120$^\circ $ angles with the neighboring spins.
As 
shown in Fig.2, the ground state of such triangular 
{\it XY\/}
spins is two-fold degenerate 
according as the resulting
noncollinear spin structure is either right- or left-handed 
(chiral degeneracy). A given chiral state cannot be transformed into the state
with the opposite chirality via any global spin  rotation
in the {\it XY\/}-spin space, global spin {\it reflection\/} being
required to achieve this.
One may assign a {\it chirality\/} 
+ and $-$ to each of these two ground states. 
In other words, the ground-state
manifold of the frustrated {\it XY\/} magnets 
possess a hidden Ising-like discrete degeneracy, chiral degeneracy,
in addition to a continuous degeneracy associated with 
the continuous {\it XY\/}-spin symmetry. 
The concept of  chirality was introduced into
magnetism first by Villain [1].
\par
To characterize these two chiral states, it is convenient to
introduce a scalar quantity, chirality, defined by [2]
$${\bf \kappa }_p={2 \over 3\sqrt 3}
\sum _{<ij>}^p[\vec S_i\times \vec S_j]_z
={2 \over 3\sqrt 3}\sum _{<ij>}^p(S_i^xS_j^y-S_i^yS_j^x),\eqno(1.1)$$
where the summation runs over the three directed bonds surrounding
a plaquette (triangle).
One can easily confirm that $\kappa _p$ gives $\pm 1$ for the two
spin configurations depicted in Fig.2. Note that the 
chirality defined by (1.1)
is a {\it pseudoscalar\/} in the sense that it is
invariant under global spin rotation [SO(2)=U(1)] 
while it changes sign
under global spin reflection [Z$_2$].
\par
In the triangular spin structure formed by the 
$n$=3-component Heisenberg spins, by contrast, 
there is no longer a discrete chiral degeneracy since
the two spin configurations in Fig.2 can now be transformed 
to each other by continuous spin rotation 
via the third dimension of the Heisenberg spin.
However, 
one can define a chirality {\it vector\/}
as an axial vector defined by [3]
$$\vec \kappa _p={2 \over 3\sqrt 3}
\sum _{<ij>}^p\vec S_i\times  \vec S_j.\eqno(1.2)$$
\par
The situation described above is essentially the same also in
the 2D triangular and 3D 
stacked-triangular antiferromagnets. 
In the ordered state, the sublattice-magnetization vector on each 
sublattice (triangular layer consists
of three interpenetrating triangular sublattices) 
cant with each other making an angle equal to 120$^\circ $.
In the case of {\it XY\/} spins, such triangular structure gives 
rise to the chiral degeneracy
as shown in Fig.3.

Similar chiral degeneracy is also  realized in other types of canted
magnets such as helimagnets (spiral magnets), in which
right- and left-handed helices 
as illustrated in Fig.4  are 
energetically degenerate.
\par\bigskip
\noindent
(c) Short history of research
\par\medskip
Historically, studies on the critical properties of  canted or 
noncollinear magnets 
was initiated more than 20 years ago for 
rare-earth helimagnets Ho, Dy and Tb. In 1976, Bak and Mukamel
analyzed theoretically 
the critical properties of the paramagnetic-helimagnetic 
transition of easy-plane-type ({\it XY\/}-like) helimagnets
Ho, Dy and Tb [4].
Bak and Mukamel derived an effective Hamiltonian called 
Landau-Ginzburg-Wilson (LGW) Hamiltonian
appropriate for the {\it XY\/} ($n=2$) helimagnet and performed a
renormalization-group (RG) $\epsilon =4-d$ expansion analysis. They
found a stable $O(4)$-like fixed point and claimed that Ho, Dy and Tb
should exhibit a continuous transition characterized by the standard
$O(4)$-like exponents $\alpha \simeq -0.17$, $\beta \simeq 0.39$, 
$\gamma \simeq 1.39$ and $\nu \simeq 0.70$. 
Note that the predicted singularity is
weaker than that of 
the unfrustrated collinear {\it XY\/} magnet; namely,
$\alpha $ is more negative while  $\beta $, $\gamma $ and $\nu$ 
are larger. Similar $\epsilon $-expansion analysis with  
interest in the commensurability effect on the helical transition 
was also made by
Garel and Pheuty [5], who found, for the case of {\it XY\/}  ($n=2$)
spins, the
same $O(4)$-like fixed point as obtained by Bak and Mukamel.
\par
Meanwhile, experiments on rare-earth helimagnets Ho, Dy and Tb
gave somewhat inconclusive results. Some of these experiments,
especially  neutron-diffraction measurements 
for Ho [6], supported  the predicted $O(4)$ behavior,
while some other experiments, such as  specific-heat
measurements  for Dy [7], 
M\"osbauer measurements  for Dy [8], 
and neutron-diffraction measurements  for Tb [9],
yielded exponents  significantly different 
from the  $O(4)$ values.
\par
A few years later,  Barak and
Walker reanalyzed the RG calculation by Bak and Mukamel, and found
that the $O(4)$-like fixed point found by them was actually located
in the region of the parameter space representing the 
{\it collinear\/} spin-density-wave (SDW) order,  not
the noncollinear helical order [10].
Since no stable fixed was found 
in the appropriate region in
the parameter space, Barak and Walker 
concluded that the paramagnetic-helimagnetic
transition of Ho, Dy and Tb should be first order.
Although most of the experimental works 
on Ho, Dy and Tb done so far have reported a
continuous transition, a few authors suggested that the transition
of Ho and Dy might actually be weakly first order [11,12]. 
In fact, experimental situation concerning the critical properties of
these rare-earth helimagnets has remained confused for years now, 
in the sense that different authors reported significantly different
exponent values, or even different order of the transition, 
{\it for the same exponent of the same material\/}. 
For example, the reported values of the exponent $\beta $
are scattered from 0.21 (Tb; X-ray) [13], 0.23
(Tb; neutron) [14], 0.25(Tb; neutron) [9], 0.3(Ho; neutron) [15], 
0.335(Dy; M\"osbauer) [8], 
0.37(Ho; X-ray) [16], 0.38(Dy; neutron) [17], 
0.39(Ho; neutron) [18] to 0.39(Dy; neutron) [18].
\par
In 1985-6, 
first theoretical analysis of the
critical properties of  stacked-triangular antiferromagnets
was made by the present author 
for both cases of {\it XY\/} and Heisenberg spins [19,20]. 
By means of a
symmetry analysis and Monte Carlo simulations, it was claimed that,
due to its chiral degrees of freedom,
phase transition of these stacked-triangular antiferromagnets
might be novel, possibly belonging to a new universality,
called the {\it chiral universality class\/},
different from the standard $O(n)$ Wilson-Fisher
universality class. 
The critical singularity observed in 
Monte Carlo simulations was
stronger than that of the unfrustrated collinear {\it XY\/}
and Heisenberg magnets, opposite to the Bak and Mukamel's 
$O(4)$ prediction. Indeed, the exponent values determined by 
Monte Carlo simulations were 
$\alpha = 0.34\pm 0.06$, $\beta = 0.253\pm 0.01$, $\gamma
= 1.13\pm 0.05$ and $\nu = 0.54\pm 0.02$ for the {\it XY\/} case, and
$\alpha = 0.24\pm 0.08$, $\beta = 0.30\pm 0.02$, $\gamma
= 1.17\pm 0.07$ and $\nu = 0.59\pm 0.02$ for the Heisenberg case [21].
It was  predicted that such  novel 
critical behavior should be
observed in the stacked-triangular {\it XY\/} antiferromagnet
CsMnBr$_3$ (n=2 chiral universality) [20,22], and in the 
stacked-triangular Heisenberg antiferromagnets VCl$_2$ and VBr$_2$
(n=3 chiral universality) [19,22], 
while helimagnets such as Ho, Dy and Tb
were also argued to exhibit the same novel 
$n=2$ chiral critical behavior asymptotically [20,22]. 
RG $\epsilon =4-d$ and $1/n$ expansion analyses were also made
by the  author, 
and a new fixed point describing the noncollinear criticality
was identified [23].
\par
Stimulated by this theoretical prediction, 
several experiments were  subsequently made  on the critical
properties of stacked-triangular antiferromagnets
CsMnBr$_3$, VCl$_2$ and VBr$_2$.
The first experimental measurements of the critical properties of the
stacked-triangular 
{\it XY\/} antiferromagnet CsMnBr$_3$ were performed
by the two groups, {\it i.e.\/,} 
neutron-scattering measurements 
by the McMaster group (Mason, Gaulin and Collins) [24] and the one
by the Japanese group (Ajiro, Kadowaki and coworkers) [25].
The results of
these two independent
measurements were consistent with each other and yielded
the exponent values close to the predicted  values, 
giving some support to the
chiral-universality scenario.
Since then, further measurements have been performed on
CsMnBr$_3$, including  
high-precision specific-heat measurements by
the Santa Cruz group (Wang, Belanger and Gaulin) [26]
and the one by the Karlsruhe group
(Deutschmann, Wosnitza, von L\"ohneysen and Kremer) [27]. 
For the stacked-triangular 
Heisenberg antiferromagnets VCl$_2$ and VBr$_2$, 
following  the first specific-heat measurements 
by Takeda and coworkers [28],
both neutron-scattering [29] and specific-heat [30] measurements were 
performed. Most of the obtained exponents and the specific-heat 
amplitude ratio were in
reasonable agreement with the predicted 
values. 
\par
By contrast,
a more conservative view
was proposed by Azaria, Delamotte and Jolicoeur a few years later
[31,32].
These authors studied a certain nonlinear sigma model expected to
describe
the Heisenberg ($n=3$) noncollinear or canted magnets based on the
RG $\epsilon =d-2$ expansion technique, and found a stable fixed point
which was nothing but the standard  $O(4)$ Wilson-Fisher fixed point.
These authors then suggested that the magnetic phase transition of
noncollinear magnets, including both 
stacked-triangular antiferromagnets and helimagnets, 
might be of standard $O(4)$ 
universality. The O(4) fixed point found there for the
Heisenberg spins is different in nature from the
$O(4)$-like fixed point 
found by Bak and Mukamel for the {\it XY\/} spins [4]:
The former $O(4)$ fixed point 
has no counterpart in the $\epsilon =4-d$ expansion.
Azaria {\it et al\/} further speculated that the noncollinear 
transition could be either first order or mean-field tricritical
depending on the microscopic properties of the system.
\par
One useful method to directly 
test those theoretical predictions is a
Monte Carlo simulation on a simple spin model.
Following the first Monte Carlo study on the {\it XY\/} and
Heisenberg stacked-triangular antiferromagnets [19-21],
extensive Monte Carlo simulations have been performed by 
several different groups, including Saclay group
(Bhattacharya, Billore, Lacaze and Jolicoeur; Heisenberg) [33], 
Cergy group (Loison, Boubcheur and Diep; 
{\it XY\/} [34] and Heisenberg [35]), 
and by Sherbrooke group (Mailhot, Plumer and
Caill\'e; {\it XY\/} [36] and Heisenberg [37]). 
In the numerical sense, the reported 
results  agreed 
with each other and with the earlier simulation of Ref. 21,  
except for a small difference left in some exponents of
the {\it XY\/} system. More specifically, 
in the Heisenberg case, 
the results support the 
chiral-universality scenario in the sense that  a continuous
transition characterized by the novel exponents 
were observed in common.
In particular, one may now rule out the possibility of the
standard $O(4)$ critical behavior and of the mean-field tricritical
behavior predicted  by
Azaria {\it et al\/}.
In the {\it XY\/} case, the results are again
consistent with the chiral-universality scenario, but 
inconsistent with the $O(4)$-like behavior predicted 
by Bak and Mukamel.
Meanwhile, since the
exponent values predicted for the $n=2$ chiral-universality are
not much different from the mean-field tricritical values 
$\alpha =0.5$, $\beta =0.25$ and $\gamma =1$, 
some authors interpreted their Monte Carlo
results on the {\it XY\/} model
in favor of the mean-field tricritical behavior rather
than  the chiral universality [36].
One should also bear in mind
that the possibility
of a weak first-order transition may not completely  be 
ruled out
from numerical simulations for finite lattices.

Important progress was also made in the study of the magnetic phase
diagram and the  multicritical behavior of stacked-triangular
antiferromagnets under external magnetic fields.
In particular,  magnetic phase diagram with a novel multicritical
point,  different from those of the standard unfrustrated
antiferromagnets, was observed by Johnson, Rayne and Friedberg for
weakly Ising-like stacked-triangular antiferromagnet CsNiCl$_3$ by 
susceptibility measurements [38]. For the 
stacked-triangular {\it XY\/} 
antiferromagnet CsMnBr$_3$, Gaulin, Mason, Collins
and Larese revealed by neutron-scattering measurements that
the zero-field transition point corresponds to a tetracritical point in
the magnetic field -- temperature  phase diagram [39]. 
These novel 
critical and multicritical properties of stacked-triangular
antiferromagnets under external fields were theoretically investigated 
by Kawamura, Caill\'e and Plumer  within a scaling 
theory based on the chiral-universality scenario, and several
prediction were made [40]. To test these scaling predictions,
further experiments were performed in turn, which revealed
features of the noncollinear transitions under external fields.
\par\bigskip
\noindent
(d) Outline of the article
\par\medskip
In the following sections, I wish to review 
in more detail these theoretical and
experimental studies concerning the critical properties of
noncollinear or canted  magnets [41-44].
In \S2, I will explain several typical magnetic
materials exhibiting the
noncollinear spin order, and introduce simple
spin models used in describing these noncollinear
transitions. The LGW Hamiltonian
appropriate for the noncollinear 
transitions is also introduced.
In \S3, an intuitive symmetry argument 
is given on the basis of the notion of the order-parameter space.
Symmetry properties of the LGW Hamiltonian
is also examined. Analysis of topological defects  in the 
noncollinearly-ordered state is given, and the nature of topological
phase transitions mediated by the topological defects
is briefly discussed.
Section 4 is devoted to the RG analyses of the noncollinear
transitions,
including $\epsilon =4-d$ expansion, $1/n$ expansion and $\epsilon
=d-2$ expansion. After presenting the results of these RG 
calculations,
several different theoretical proposals are explained and discussed.
In \S 5, the results of Monte Carlo simulations on the
critical properties of {\it XY\/} and Heisenberg
stacked-triangular antiferromagnets and of several related models
are presented.
In \S 6, 
experimental results on the critical properties of both 
stacked-triangular antiferromagnets and helimagnets are 
reviewed.
A possible experimental method to measure the chirality
is mentioned.
The phase transition of stacked-triangular 
antiferromagnets  under external magnetic 
fields is reviewed in \S 7, 
with particular emphasis on its phase diagram and
novel multicritical behavior.
Finally, in \S 8, 
I summarize the present status of the study, and discuss future 
problems.
\par
\bigskip
\bigskip
\noindent
{\bf \S 2. Materials and Models}
\par
\medskip
In this section, I introduce typical materials and model systems
which have been used in the study of noncollinear phase transitions.
These include both (a) stacked-triangular antiferromagnets and (b)
helimagnets.
\par\bigskip
\noindent
(a) Stacked-triangular antiferromagnets
\par\medskip
In stacked-triangular antiferromagnets, magnetic ions are located at
each site of a 
three-dimensional stacked-triangular (simple hexagonal)
lattice.
Magnetic ions  interact  antiferromagnetically
in the triangular layer, which
causes the geometry-induced frustration.
Most extensively studied stacked-triangular antiferromagnets 
are ABX$_3$-type compounds, 
A being elements such as Cs and Rb, 
B being magnetic ions
such as Mn, Cu, Ni, Co, and C being halogens such as Cl, Br and I
[42,44].
While these materials are magnetically
quasi-one-dimensional, it has been established 
that most of them exhibit a magnetic transition into a 
three-dimensionally ordered state  at low
temperatures with sharp magnetic Bragg peaks.
There is a rich variety of materials 
depending on the combination of
the constituent ions, A, B and C [44]. 

Crucial to the nature of phase transition is the type
of magnetic anisotropy. Some of these compounds 
are Ising-like with easy-axis-type (or axial) anisotropy, 
some are XY-like with easy-plane-type (or planar)
anisotropy, and others are Heisenberg-like with negligibly small
anisotropy. In zero field, 
the noncollinear criticality is 
realized in the {\it XY\/} and Heisenberg systems, which include
CsMnBr$_3$, CsVBr$_3$ (XY), CsVCl$_3$ and
RbNiCl$_3$ (nearly Heisenberg) {\it etc\/}.
By contrast, the Ising-like axial magnets including 
CsNiCl$_3$, CsNiBr$_3$,  and CsMnI$_3$  often  exhibit two
successive phase transitions in zero field
with the collinearly-ordered intermediate phase.
If an external field of appropriate intensity is applied along an
easy-axis, however, 
a direct transition from the
paramagnetic state to the noncollinearly-ordered 
state becomes possible.
Such  transition in an external field  is characterized by 
the nontrivial chirality,
and
will  also be discussed later in \S 6 and \S 7.
\par
Quasi-two-dimensional realization of stacked-triangular
antiferromagnets may be vanadium compounds VX$_2$ with X=Cl and Br.
VX$_2$ are nearly isotropic (Heisenberg-like) magnets with weak
Ising-like anisotropy.
While VX$_2$ exhibits two successive
transitions at two distinct but mutually 
close
temperatures due to the weak easy-axis-type anisotropy 
($T_{N1}\simeq 35.88K$ and $T_{N2}\simeq 35.80K$ in case
of VCl$_2$ [29]), it is expected to behave 
as an isotropic Heisenberg system 
except   close to $T_{N1}$ or $T_{N2}$.
\par
Since our interest is on the noncollinear criticality,
we will mainly be concerned in this article  
with vector spin systems,
including both $n=2$-component {\it XY\/} 
and $n=3$-component Heisenberg spin models.
A simple vector-spin Hamiltonian often used in modeling such
stacked-triangular antiferromagnets may be given by
$${\cal H}=-J\sum _{<ij>} \vec S_i\cdot \vec S_j
-J'\sum _{<<ij>>}\vec S_i\cdot \vec S_j,\eqno (2.1)$$
where $\vec S_i=(S_i^{(1)},S_i^{(2)},\cdots ,S_i^{(n)})$ 
is an $n$-component unit vector with $\mid \vec S_i
\mid =1$ located at the $i$-th site of a stacked-triangular lattice, 
while
$J<0$ and $J'$ represent the intraplane and 
interplane nearest-neighbor couplings.
The first sum is taken over all nearest-neighbor pairs
in the triangular layer, while the second sum is taken over
all nearest-neighbor pairs along the chain direction orthogonal to
the triangular layer.
\par 
\bigskip
\noindent
(b) helimagnets
\par\medskip
The second class of noncollinear magnets 
is a helimagnet or spiral magnet. Examples are $\beta $-MnO$_2$ and 
rare-earth metals Ho, Dy and Tb.
Rare-earth helimagnets Ho, Dy and Tb crystallize into 
the hexagonal-closed-packed (hcp) structure,
and form magnetic spiral 
along the $c$-axis below $T_N$ 
with the moments lying in the basal plane.
The interaction between magnetic moments is the 
long-range Ruderman-Kittel-Kasuya-Yoshida (RKKY) interaction
which falls off as $1/r^3$ and oscillates in sign
with distance $r$. The oscillating nature of the RKKY interaction
leads to the frustration between the near-neighbor and further-neighbor
interactions which stabilizes the noncollinear helical spin structure.
\par
A simple model Hamiltonian which gives rise to a spiral structure
is the axial-next-nearest-neighbor {\it XY\/} or Heisenberg
model on a simple cubic lattice,
with the ferromagnetic (or
antiferromagnetic) nearest-neighbor interaction in all directions 
and the antiferromagnetic next-nearest-neighbor interaction 
along one particular direction, say
$x$ direction.
The Hamiltonian may be written as
$${\cal H}=-J_1\sum _{<ij>}\vec S_i\cdot \vec S_j
-J_2\sum _{<<ij>>}\vec S_i\cdot \vec S_j, \eqno (2.2)$$
where the first sum
is taken over all nearest-neighbor pairs on the lattice while
the second sum is taken over next-nearest-neighbor pairs
along the $x$-direction. 
The competition between the nearest-neighbor interaction $J_1$
and the antiferromagnetic axial next-nearest-neighbor interactions 
$J_2<0$ gives rise to a magnetic spiral along the $x$ direction
when the value of $\mid J_2 /J_1\mid $ exceeds a certain critical value.

One difference of such spiral structure 
from the noncollinear structure in the
stacked-triangular antiferromagnet
is that
the pitch of the helix is generally
incommensurate with the underlying lattice, in contrast to the
120$^\circ $ spin structure which is always commensurate with
the underlying lattice.
(In fact, one can generate the incommensurate spin structure even 
in stacked-triangular antiferromagnets, {\it e.g.\/}, 
by breaking the equivalence 
of the intraplane couplings [45,46]. 
This case might have some relevance
to the incommensurate spin order in RbMnBr$_3$ 
as will be discussed in \S 6.)
\par
\bigskip
\noindent
(c) Landau-Ginzburg-Wilson (LGW) Hamiltonian
\par\medskip
The spin Hamiltonians (2.1) and (2.2)  have been written
in terms of the spin variables of fixed length, $\mid \vec
S_i\mid =1$.
In some of 
the RG analyses such as  $\epsilon =4-d$ or 
$1/n$ expansions, an alternative form of Hamiltonian 
written in terms of 
spin-variables of unconstrained length is often used.
It is given in the form of an expansion in order-parameter
fields (critical modes), and is called
the Landau-Ginzburg-Wilson (LGW) Hamiltonian. 
In the case of standard ferromagnets or unfrustrated collinear
antiferromagnets, an appropriate LGW Hamiltonian is the so-called
$\vec \phi ^4$ model whose Hamiltonian density is given  by
$${\cal H}_{{\rm LGW}}={1\over 2}
[(\bigtriangledown {\bf \vec \phi})^2+
r{\bf \vec \phi}^2+u{\bf \vec \phi}^4],
\eqno(2.3)$$
where  $n$-component vector field ${\bf \vec \phi} 
=(\phi_1,
\phi_2,\cdots ,\phi_n)$ represents near-critical mode
around an instability point.
In unfrustrated ferromagnets or antiferromagnets,
the instability occurs only at one point in the 
wavevector space, as shown in Figs.5a and b. Therefore, 
single $n$-vector field $\vec \phi $
is enough to describe the phase transition.
\par
By contrast, in the case of noncollinear or canted magnets
such as stacked-triangualr antiferromagnets or helimagnets,
the instability occurs simultaneously at two distinct points in
the  wavevector space. Therefore,  
two equivalent but distinct $n$-component
vector fields are necessary to describe the associated phase
transition. The situation is illustrated in Figs. 5c and 5d
for the cases of stacked-triangular antiferromagnets
and helimagnets, respectively.
These two instability modes may be taken as
the Fourier modes at $\pm  \vec Q$, where $\vec Q=
(4\pi /3,0,0, \cdots ,0)$ 
for the case of stacked-triangualr antiferromagnets, 
and $\vec Q=(2\pi /\lambda, 0,0,\cdots ,0)$ for the case of
helimagnets, 
$\lambda $ being the pitch of the helix.
It is convenient for later use to extend the model to general $d$
spatial dimensions. In the case of 
stacked-triangular antiferromagnets, 
the lattice is then regarded as two-dimensional triangular
layers stacked in hypercubic fashion 
along the remaining $d-2$ directions, while
in the case of helimagnets, the competing second-neighbor 
interaction is assumed to work only along the first direction
in $d$ dimensions, along which the helix is formed.
\par
One can derive the soft-spin LGW Hamiltonian 
starting from the
microscopic hard-spin Hamiltonian (2.1) or (2.2) by a series of
transformations [23].
By softening the fixed-length spin condition, 
Fourier transforming, and retaining only near critical
models, one obtains
$${\cal H}_{{\rm LGW}}={1\over 2}[(\bigtriangledown \vec a)^2+
(\bigtriangledown \vec b)^2+r(\vec a^2+\vec b^2)
+u(\vec a^2+\vec b^2)^2+v\{(\vec a\cdot \vec b)^2-
\vec a^2\vec b^2\}],
\eqno(2.4)$$
where $\vec a$ and 
$\vec b$ are $n$-component vector fields
representing the cosine and sine components associated with the
noncollinear spin structure 
at wavevectors $\pm \vec Q$
via, 
$$\vec S(\vec r)=\vec a(\vec r)\cos (\vec Q\cdot \vec r)+
\vec b(\vec r)\sin (\vec Q\cdot \vec r).\eqno(2.5)$$
In order that the spin structure (2.5) really
represents the noncollinear order,
the $\vec a$ and $\vec  b$ fields must be orthogonal with
each other. This requires that the
quartic coupling $v$ in the LGW Hamiltonian (2.4) should be positive.
If $v$ is negative, on the other hand, 
the spin structure given by (2.5) represents the
collinearly-ordered SDW state (or the sinusoidal state). The LGW
Hamiltonian (2.4) forms a basis of the following 
RG $\epsilon =4-d$ and $1/n$
expansion analysis.
\par
In the particular case of {\it XY\/}  ($n=2$) spins, 
one can transform (2.4) into
a different form [23],
$${\cal H}_{{\rm LGW}}={1\over 2}[(\bigtriangledown \vec A)^2+
(\bigtriangledown \vec B)^2+r(\vec A^2+\vec B^2)
+(u-{1\over 4}v)(\vec A^4+\vec B^4)+2(u+{1\over 4}v)\vec A^2
\vec B^2], 
\eqno(2.6)$$
where $\vec A$ and $\vec B$ are two-component fields defined by
$$A_x=(a_x+b_y)/\sqrt 2, \ \ \ B_x=(a_y+b_x)/\sqrt 2,  $$
$$A_y=(a_y-b_x)/\sqrt 2, \ \ \ B_y=(-a_x+b_y)/\sqrt 2.  \eqno(2.7)$$
The RG analysis of Ref. 4
was performed on the basis of the form (2.6), rather than (2.4).
From (2.6), it is easy to see that, in the  case of $n=2$, 
the model reduces to two decoupled {\it XY\/} models on the
special manifold $v=-4u$. 
Note that this manifold lies in the sinusoidal region, $v<0$.
\par
Essentially the same LGW Hamiltonian 
has also been used  in
other problems such as 
the phase transition of the dipole-locked A phase of
helium three [47,48], the superconducting phase transition of
the heavy fermion superconductor UPt$_3$ [49], and the quantum phase 
transition of certain Josephson junction array in a magnetic
field [50].
\par
\bigskip
\bigskip
\noindent
{\bf \S 3. Symmetry}
\par
\medskip
\noindent
(a) Symmetry of the ordered state
\par
\medskip
Because of its nontrivial chiral degrees of freedom,
symmetry of the ordered state of frustrated noncollinear magnets
differs from that of unfrustrated collinear magnets.
Let us consider, for example, the case of the
$n$=3-component Heisenberg spins. In the unfrustrated collinear case,
spins align parallel or antiparallel with each other forming the
collinear ground state. One can see
that such a ground state
is invariant under the global spin 
rotation around the magnetization (or the sublattice
magnetization) axis. In the frustrated noncollinear case, 
by contrast, the
120$^\circ $  spin structure does not have such an invariance.
Therefore, 
symmetries of the ordered states are clearly different in 
the collinear and the noncollinear cases. Obviously, the conventional
index $n$, the number of the spin components, is inadequate to
distinguish between such differences in the symmetry of the ordered
states. 

In order to characterize the relevant symmetry, it is 
convenient to introduce the notion of order-parameter space, 
which is a
topological space isomorphic to the set of ordered states [51].
In the collinear case, the order-parameter space $V$ may be
represented by a single arrow in the three-dimensional spin space
and is isomorphic to the two-dimensional sphere $S_2$ (the surface of
a ball in Euclidean three-space). In the noncollinear case,
the order-parameter space cannot be represented by a single arrow.
Instead, 
additional structure caused by noncollinear
alignment of  spins leads to an order-parameter space isomorphic
to the three-dimensional rotation group $SO(3)$, or equivalently,
to the projective space $P_3$ [3]. 
In the  collinear case,
rotation invariance around the magnetization axis
reduces the order-parameter space to $V=SO(3)/SO(2)=S_2$. 
\par
In the case of the $n$=2-component {\it XY\/} spins,
the order-parameter space of unfrustrated collinear 
systems is $V=S_1=SO(2)$,
while that of frustrated noncollinear systems is 
$V=Z_2\times S_1=Z_2\times SO(2)=O(2)$ where $Z_2$ pertains to the
aforementioned twofold chiral degeneracy 
while $S_1=SO(2)$ pertains
the rotation symmetry of the original {\it XY\/} spins.
\par
Order-parameter space may also be defined as a topological space
obtained by dividing the whole symmetry group of the 
Hamiltonian, which we assume to be $O(n)$,
by the subgroup which keeps the ordered state 
(symmetry-broken state) unchanged [51].
With use of this definition, one can easily generalize
the argument to
the general $n\geq $2-component vector spins.
In the unfrustrated collinear case, the invariant 
subgroup turns out to be 
$O(n-1)$, consisting of
the rotation around the magnetization axis. This leads to 
the associated order-parameter space isomorphic to the
$(n-1)$-dimensional hypersphere, 
$V=O(n)/O(n-1)=SO(n)/SO(n-1)=S_{n-1}$. In the particular
cases of $n$=2 or 3,
this simply reproduces the
results mentioned above.
\par
In the frustrated noncollinear case, 
if one notes that the 120$^\circ $
spin structure spans the two-dimensional subspace in $n$-dimensional
spin space,
one may see that 
the invariant subgroup is $O(n-2)$ 
rather than $O(n-1)$. Thus, the order-parameter space 
for the $n\geq 2$-component
noncollinear systems is isomorphic to the Stiefel manifold,
$V=O(n)/O(n-2)$ [41]. In the $n$=2 case, 
it reduces to $V=O(2)$ since $O(0)=1$, whereas in the $n=3$ case, 
it reduces to $V=SO(3)$ since $O(1)=Z_2$.
\par
Thus, the difference in the symmetry of the ordered states can be
described in topological terms as the difference in the associated
order-parameter spaces. Since the symmetry of the ordered state is a
crucial ingredient of the corresponding disordering phase transition,
this observation strongly suggests that the frustrated 
noncollinear magnets might exhibit a novel phase 
transition, possibly belonging to a new universality class [19,20]. 
Of course, another  possibility might be that these noncollinear
magnets exhibit a 
first-order transition. One cannot even rule out the possibility
that the symmetry is dynamically restored at the transition, 
and the noncollinear transition is of  conventional 
Wilson-Fisher universality class. 
In order to determine
which of the above possibilities is actually the case,
more detailed analysis 
is needed.
Still, the fact that 
one obtains for frustrated noncollinear magnets 
the order-parameter space  
different from that for the  unfrustrated collinear
magnets gives a hint
that something new
may happen in the noncollinear transitions.
\par\bigskip
\noindent
(b) Symmetry of the LGW Hamiltonian
\par
\medskip
Next, let us examine the symmetry property of the
LGW Hamiltonian of noncollinear magnets with $n$-component spins,
eq.(2.4).
The LGW Hamiltonian is invariant under the following two
symmetry transformations;
that is
(i) $O(n)$ spin rotation, $\vec a'=R\vec  a$, $\vec b'=R\vec b$
with $R\in O(n)$, as well as (ii) $O(2)$ phase rotation,
$\vec a'=\cos \theta \vec a-\sin \theta \vec  b$,
$\vec b'=\pm (\sin \theta \vec a+\cos \theta \vec b$) [23].
The latter invariance arises from the arbitrariness in choosing
the phase and the handedness of the two basis vectors.
\par

Conversely, the symmetry requirements (i) and (ii) fully determine the
form of the Hamiltonian up to quartic order in the fields $\vec a$ 
and
$\vec b$ as given in (2.4).
One may easily see that this $O(n)\times O(2)$ symmetry
of the LGW Hamiltonian just
corresponds to the aforementioned order-parameter space
$V=O(n)/O(n-2)$.
\par

In the case of $n=2$, and in this  case only, 
the LGW Hamiltonian
(2.4) has a discrete symmetry independent of the above
$O(n)\times O(2)$ symmetry. This corresponds to the permutation
of the field variables, (iii) \ ($a_x'=a_x,\ a_y'=b_x,\ b_x'=a_y,
\ b_y'=b_y$) or ($a_x'=b_y,\ a_y'=a_y,\ b_x'=b_x,
\ b_y'=a_x$). 
\par\bigskip
\noindent
(c) Classification of topological defects
\par
\medskip
One property
which can be determined solely from the topological
considerations is the classification of 
topological
defects in the ordered state.
Although we leave the details of the method to Ref.51,
the point is that
one can obtain all possible topological 
defects together with their `topological
quantum number' 
from the knowledge of its order-parameter space $V$
by examining its $r\/$-th homotopy group, 
$\Pi _r(V)$.
\par
Topological defects play an essential role
in the phase transition of  two-dimensional systems.
Many two-dimensional phase transitions,
such as the Kosterlitz-Thouless transition,
are known to be `defect mediated' [52]. 
Classification of topological defects 
in the collinear and noncollinear $d=2$-dimensional 
magnets   is given in Table I for both cases of
{\it XY\/} ($n=2$) and Heisenberg ($n=3$) spins [3].

The noncollinear 2D {\it  XY\/} systems,
such as the triangular-lattice {\it XY\/} antiferromagnets and the
Josephson-junction arrays in a magnetic field, 
possess the standard
Kosterlitz-Thouless-type vortex characterized by the
integral topological quantum number $Z$ 
as well as the chiral domain wall
characterized by the two-valued topological quantum number $Z_2$.
The vortex (point defect) 
concerns  the continuous {\it XY\/} degrees of 
freedom via the relation ($\Pi _1(S_1)=Z$), while
the domain wall (line defect)
concerns  the discrete chiral degrees of freedom
via the relation ($\Pi _0(Z_2)=Z_2$).
Since earlier MC works on the 
triangular {\it XY\/} antiferromagnet by Miyashita and Shiba [2]
and by Lee {\it et al\/} [53], 
and the one on the Josephson-junction array
by Teitel and Jayaprakash [54],
many numerical
works have been made with  interest in how these two degrees of
freedom order. While the existence of a phase transition
with a sharp specific-heat anomaly 
driven  by the appearance of the
chiral long-range order has been established, the question
whether the spin and the chirality order at the same temperature,
or at two close but distinct temperatures, 
still remains somewhat controversial [55,56].

As was first observed by Kawamura and Miyashita [3], 
the noncollinear Heisenberg magnets, such as the
triangular Heisenberg antiferromagnet, possess a peculiar vortex
characterized by its quantum number $Z_2$ ($Z_2$-vortex), 
different in nature from 
the standard $Z$-vortex of the {\it XY\/} magnets. Although it
is generally believed that the two-dimensional Heisenberg model does
not exhibit any phase transition at finite temperature [57],
possible existence of a novel topological phase
transition mediated by these $Z_2$ vortices 
was suggested by Kawamura and Miyashita  in the two-dimensional
triangular Heisenberg antiferromagnet [3]. The predicted
low-temperature phase is an exotic spin-liquid phase where
the two-point spin correlation decays exponentially and the spin
correlation length remains finite. A quantity 
called vorticity modulus, characterizing such
exotic vortex order 
not accompanying the conventional spin order, 
was proposed and calculated [58,59]. 
\par
In three spatial dimensions,  our main concern here,
point defects in two dimensions appear as line defects.
Hence, the noncollinear {\it XY\/} magnets in $d=3$ dimensions
possess  
$Z$-vortex lines in addition to  the $Z_2$ chiral domain walls, 
while the noncollinear Heisenberg 
magnets possess   $Z_2$-vortex lines.  
Although it is possible and  enlightening 
to understand the nature of the three-dimensional transitions 
also
as defected-mediated [60], 
we follow more standard theoretical approaches in this
article in which these topological defects do not show up in an
explicit way. 
\par\bigskip\bigskip
\noindent
{\bf \S 4. Theoretical analysis of critical properties I 
--- renormalization-group analysis}
\par
\medskip
In this section, I will review the theoretical analysis 
of the critical properties of noncollinear transitions
based on several
renormalization-group (RG) methods in some detail, 
including  $\epsilon =4-d$
expansion, $1/n$ expansion and $\epsilon =d-2$ expansion.
\par
\medskip
\noindent
(a) Mean-field  approximation
\par
\medskip
Standard RG calculations such as 
$\epsilon =4-d$ and $1/n$ expansions are generally performed
based on the soft-spin LGW Hamiltonian.  Before
entering into the RG analysis, 
it may be instructive here to summarize the results of the
standard mean-field approximation applied to the LGW Hamiltonian,
eq.(2.4) [23].
\par
When the quartic coupling constant 
$v$ is positive and satisfies the inequality $v<4u$, 
a continuous transition takes place at $r=0$ 
between the paramagnetic  and the noncollinear states 
characterized by
$$\mid \vec a\mid ^2=\mid \vec b\mid ^2
=-r/(4u-v),\ \ \ \ \vec a \perp \vec b \ \ \ \ \ \  
(0<v<4u), \eqno (4.1a)$$
When $v$ is negative, by contrast,
there is a continuous transition at $r=0$
between the paramagnetic  and the collinearly-ordered
sinusoidal states characterized by
$$\mid \vec a\mid ^2+\mid \vec b\mid ^2
=-r/2u,\ \ \ \ \ 
\ \ \vec a \parallel \vec b \ \  \ \ \ 
(v<0). \eqno (4.1b)$$
Note that, in the sinusoidal case, 
the relative magnitude of $\vec a$
and $\vec b$ is not determined: This corresponds physically to the 
sliding degree of freedom of the spin-density wave.

Stability of the
free energy requires 
the condition 
$$u>0,\ \ v<4u.\eqno (4.2)$$
When $u<0$ or $v>4u$,  higher-order
(sixth-order) term 
is necessary 
to stabilize the free energy, and the transition in such a case 
generally becomes
{\it first order}. The mean-field
phase diagram in the $u$-$v$ plane is summarized in Fig.6.
Continuous transitions are
characterized by the standard mean-field exponents, 
$\alpha =0$, $\beta=1/2$ and $\gamma =1$ {\it etc.}, 
while 
the mean-field tricritical exponents 
$\alpha =1/2$, $\beta=1/4$ and $\gamma =1$ {\it etc.} 
are realized along the stability
boundary $v=4u$. 
Of course, fluctuations generally change these 
conclusions as we shall see below.
\medskip
\noindent
(b) $\epsilon =4-d$ expansion
\par
\medskip
In this subsection, I will review the 
RG $\epsilon =4-d$ expansion
results for the noncollinear transition.
Earlier attempts were made for  {\it XY\/} ($n=2$) helimagnets 
to $O(\epsilon ^2$) 
by Bak and Mukamel [4], 
and later by Barak and Walker [10],
with  interest in
the paramagnetic-helimagnetic transition 
of rare-earth metals Ho, Dy Tb. 
Similar $O(\epsilon ^2)$ analysis for general $n$-component
helimagnets was made by Garel and Pheuty with  interest in 
the possible
commensurability 
effect on the helical transition [5], and 
by Jones, Love and Moore [47] and by
Bailin, Love and Moore [48] 
in the context of the superfluidity transition of 
helium three. Fuller analysis
in light of  possible 
new universality class
was  made 
by the present author [23]. 
More recently, higher-order calculation to $O(\epsilon ^3)$ was made
by Antonenko, Sokolov and Varnashev [61].
Since the obtained results were sometimes interpreted in 
different ways by these authors, I will 
postpone the discussion of their physical 
implications to later subsections and will
first present
the results  based on Refs. [23] and [61].
\par
\medskip
\noindent
{\it RG flow diagram, fixed points and critical exponents\/}
\par
\medskip
Let us consider the LGW Hamiltonian for general $n$-component
noncollinear magnets, eq (2.4).
Its upper critical dimension is $d_>=4$ and a standard RG $\epsilon
=4-d$ expansion can be performed.
Near four dimensions, there are up to {\it four\/} fixed points
depending on the value of $n$. Two exist for all $n$: One is the
trivial Gaussian field point located at the origin ($u^*=v^*=0$),
which is
always unstable against both $u$ and $v$ perturbations; 
the other corresponds to the conventional isotropic $O(2n)$
Heisenberg fixed point at ($u^*>0,v^*=0$), which
is stable for sufficiently small  $n$.
To describe the remaining fixed points, we consider four distinct
regimes of relating $n$ and $d$.
\par
\medskip
\noindent
I. $n>n_{\rm I}(d)=12+4\sqrt{6}-[(36+14\sqrt 6)/3]\epsilon 
+[{137\over 150}+{91\over 300}\sqrt 6+({13\over 5}+{47\over 60}
\sqrt 6)\zeta (3)]\epsilon ^2
+O(\epsilon ^3)\simeq 21.8-23.4\epsilon +7.1\epsilon ^2+
O(\epsilon ^3)$
\par\medskip
\noindent
When $n$ is sufficiently large to meet this condition, two new
fixed points appear in the noncollinear 
region $v>0$. They may be termed
{\it chiral\/}, $C_+$, and {\it antichiral\/}, $C_-$, the former
being stable in accord with the RG flow sketched in Fig.7a.
When $n$ approaches $n_{\rm I}(d)$,
the chiral and antichiral fixed points
coalesce at a point in the upper half ($u,v$) plane and become
complex-valued for $n<n_{\rm I}(d)$.
In the sinusoidal region, $v<0$, no stable fixed points are found. 
\medskip\par\noindent
II. $n_{\rm I}(d)>n>n_{{\rm II}}(d)
=12-4\sqrt{6}-[(36-14\sqrt 6)/3]\epsilon 
+[{137\over 150}-{91\over 300}\sqrt 6+({13\over 5}-{47\over 60}\sqrt 6)
\zeta (3)]\epsilon ^2+O(\epsilon ^3)
\simeq 2.20-0.57\epsilon +0.99\epsilon ^2+O(\epsilon ^3)$
\par\medskip
\noindent
The RG flows are now as depicted in Fig.7b. Only the Gaussian and
Heisenberg fixed points are present and both are unstable.
Consequently, the transition to both noncollinear and sinusoidal
phases is expected to be first order.
\medskip\par\noindent
III. $n_{{\rm II}}(d)>n>n_{{\rm III}}(d)
=2-\epsilon 
+{5\over 24}(6\zeta (3)-1)\epsilon ^2+O(\epsilon ^3)
\simeq 2-\epsilon +1.3\epsilon ^2+O(\epsilon ^3)$
\par\medskip
\noindent
In this regime, a new pair of fixed points appear in the sinusoidal
region, $v<0$, which may be termed {\it sinusoidal\/}, $S_+$, and
{\it antisinusoidal\/}, 
$S_-$. The corresponding flows resemble those
sketched in Fig.7c.  
The fixed point $S_+$ is the fixed point identified 
by Bak and Mukamel [4], and by Garel and Pheuty [5], 
as a physical fixed point governing the {\it XY\/} ($n=2$)  
helimagnets in $d=3$. 
In the case of $n=2$,  $S_+$ coincides to $O(\epsilon )$ 
with the $O(4)$ 
fixed point, $H$, on the $v=0$ axis, while it
moves to the lower-half plane 
at higher order in $\epsilon $.
Thus, $S_+$ is the $O(4)$-like fixed point to $O(\epsilon ^2)$ in the
sense that all exponents agree with the isotropic $O(4)$ exponents,
but
it is not exactly an $O(4)$ fixed point as can be 
confirmed by the  higher-order calculation [62].
In any case,  
this Bak and Mukamel fixed point  
is located in the sinusoidal region $v<0$, and cannot be invoked to
describe the noncollinear phase transitions [10]. 
As $n\rightarrow n_{{\rm III}}(d)$, the sinusoidal
fixed point $S_+$ approaches the $v=0$ axis and, 
at $n=n_{{\rm III}}(d)$, it
meets the Heisenberg fixed point $H$ and exchanges stability with it.
In the noncollinear region $v>0$, no stable
fixed point exists.
\medskip\par\noindent
IV. $n>n_{{\rm III}}(d)$
\par\medskip
\noindent
As illustrated in Fig.7d, the unstable fixed point $S_+$ now lies 
above the $v=0$ axis. The Heisenberg fixed point $H$ is stable and
governs the critical behavior of regions of both
noncollinear and sinusoidal ordered behavior.
\par
In view of the above four cases, one can see that,
in the noncollinear region  $v>0$, the 
stable fixed point describing the
noncollinear transition
is either the chiral fixed point 
$C_+$, which is stable for sufficiently large $n$
$$n>n_{{\rm I}}(d)=21.8-23.4\epsilon +7.1\epsilon ^2+
O(\epsilon ^3),
\eqno (4.3)$$ 
or the $O(2n)$ Heisenberg fixed point $H$,
stable for sufficiently small $n<n_{{\rm III}}(d)$.
At these stable fixed points, critical exponents can be calculated
in the standard manner. The exponents at the standard Heisenberg fixed
point are well-known, 
while
the ones at the chiral fixed point are new.
To the lowest-order, the exponents $\gamma $ and $\nu $ 
at the chiral fixed point were calculated as
[23]
$$\gamma \approx 2\nu =1+{n(n^2+n+48)+(n+4)(n-3)\sqrt
{n^2-24n+48}\over 4(n^3+4n^2-24n+144)}\epsilon +O(\epsilon ^2).
\eqno(4.4)$$
These $\gamma $ and
$\nu $ are numerically smaller than the corresponding
$O(n)$ Heisenberg values.
The critical-point decay exponent to $O(\epsilon ^2)$ was 
calculated as [23]
$$\eta ={n(n^2+n+48)+(n+4)(n-3)\sqrt
{n^2-24n+48}\over 4(n^3+4n^2-24n+144)}\epsilon ^2+O(\epsilon ^3).
\eqno(4.5)$$
In the noncollinear region $v>0$, 
the facts
concerning the stable fixed points are summarized in Fig.8. 

Crucial question is 
what happens at the physically significant 
points, $\epsilon =1$ ($d=3$) with $n=2$ and 3. Unfortunately,
these are rather far from the $\epsilon \rightarrow 0$ limit, and 
thus, it is very difficult to obtain
truly definitive answer from the 
$\epsilon $ expansion with only a few terms. In fact, different
authors gave different conjectures. 
The existence of the chiral fixed point $C_+$ was first noticed
for large enough $n$ ($n>21.8$) by Moore and coworkers
in Refs. 46 and 47
in the context of helium three, while these authors claimed that
the transition in the physical case ($n=3$, $d=3$) 
was first order since
$n=3$ was significantly smaller than $21.8$.
Detailed study of the chiral fixed point, including the
$\epsilon $-expansion expression of the stability boundary $n_{\rm I}
(d)$, was 
first given in Ref.23, where it was argued in view
of the Monte Carlo results that
the chiral fixed point might remain stable down to $n=2$ or 3 in
$d=3$.  In contrast,
Antonenko,  Sokolov and Varnashev claimed based on their 
$O(\epsilon ^2)$ expression of
$n_{\rm I}(d)$ and its Borel-Pad\'e resummation  
that the transition  in $d=3$
was first order for both $n=2$ and 3 [61].

Instead of the $\epsilon =4-d$ expansion where the dimension $d$ 
is expanded in powers of $\epsilon $, one can also perform the RG
loop expansion 
directly at $d=3$. This was also done by Antonenko and Sokolov 
to three-loop order, yielding the results similar to the 
$\epsilon $-expansion calculation to the same order [63].
\par
Note also that, if one makes
the standard $\epsilon $ expansion with fixing $n$ at $n=2$ or 3 
(or any value smaller than  21.8),  
the chiral fixed point can never be seen [4].
This is 
simply because the $\epsilon $-expansion method can detect only
the type of 
fixed point which 
exists, stable or unstable, in the $\epsilon \rightarrow 0$ limit.
\par
In the special case of {\it XY\/} ($n=2$) sinusoidal ordering $v<0$,
one can give a nonperturbative
argument to identify the stable fixed point in $d=3$,
making use of the fact that the system reduces to the decoupled
{\it XY\/} models on the line $v=-4u$.
In the {\it XY\/} case, the fixed point $S_-$ is located on this
$v=-4u$ line and becomes the standard {\it XY\/} fixed point
($O(2)$ Wilson-Fisher fixed point). One can then
show based on 
nonperturbative argument that this {\it XY\/} fixed point is stable
in $d=3$ [64].
This is in contrast to the behavior 
obtained from the low-order $\epsilon $ expansion
as sketched 
in Fig.7c,
where the fixed point $S_-$
is unstable [61,62,65]. 
Unfortunately, this discrepancy between the low-order 
$\epsilon $-expansion result and the nonperturbative result
cannot be remedied
even if one goes to higher order,  say to O($\epsilon ^3$),
and makes a resummation procedure [62].
This observation gives us a warning that one should not
overtrust the answer from the $\epsilon =4-d$ expansion 
in some subtle cases, 
even when relatively higher-order calculation, say to $O(\epsilon ^3)$,
was made together with the resummation technique. 
\par
\bigskip
\noindent
{\it Chirality and other composite operators}
\par
\medskip
In this subsection, we show
how the chirality, defined in \S 1(b) as a quantity characterizing the
noncollinear spin structure, manifests itself
in the RG $\epsilon =4-d$ expansion.
As shown in \S 1(b), the chirality is a pseudoscalar in the {\it XY\/}
case and an axial vector in the Heisenberg case.
In accord with the LGW Hamiltonian (2.4),
one can also generalize the definition of the chirality for general
$n$-component spins as a second-rank antisymmetric tensor variable
defined by $\kappa _{\lambda ,\nu }=a_\lambda b_\mu - 
a_\mu b_\lambda \ (1\leq \lambda,\mu \leq n)$, which has
$n(n-1)/2$ independent components [23]. 
\par

One may define a conjugate
{chiral field\/}, $h_\kappa $, which couples to a component of the
chirality via a term $-h_\kappa \kappa _{\lambda,\nu}$ in the
LGW Hamiltonian. Application of the chiral field $h_\kappa $
reduces the original symmetry of the Hamiltonian. The noncollinear 
structure is then confined to the $(\lambda,\mu)$ plane and one
out of two senses of the helix is selected. It is thus expected
that the application of $h_\kappa $ causes a  crossover 
from the fully chiral
behavior
to the standard {\it XY\/} behavior.
\par
If there is a stable fixed for $h_\kappa =0$, 
say, a chiral fixed point,
this crossover is governed by
the chiral crossover exponent $\phi _\kappa $
associated with that fixed point.
The singular part of the free energy then has a scaling form [23],
$$f_{sing}\approx F({h\over t^\Delta}, 
{h_\kappa \over t^{\phi _\kappa }}),\eqno(4.6)$$
where $h$ is an ordering field conjugate to the order
parameter $\vec a$ or $\vec b$,
$\Delta \equiv \beta +\gamma $ is the gap exponent
(the crossover exponent associated with the ordering field),
and $t\equiv \mid (T-T_c)/T_c\mid $.
If the total chirality, $\bar \kappa=
-(\partial f/\partial h_\kappa )_{h_\kappa =0}$, and the
chiral susceptibility, $\chi _\kappa =
-(\partial ^2f/\partial h_\kappa ^2)_{h_\kappa =0}$, are characterized
by critical exponents $\beta _\kappa $ and $\gamma _\kappa $,
the above scaling gives $\beta _\kappa =2-\alpha-\phi _\kappa $ and
$\gamma _\kappa =2\phi _\kappa -(2-\alpha)$, and the chirality 
exponents satisfy the relation,
$$\alpha +2\beta _\kappa +\gamma _\kappa =2,\eqno (4.7)$$
together with the standard relation $\alpha +2\beta +\gamma =2.$
\par
In particular, in the region $n>n_{\rm I}(d)$ 
where the chiral fixed point
is stable, the chiral crossover exponent $\phi _\kappa $
has been calculated by the
$\epsilon =4-d$ expansion as [23]
$$\phi _\kappa =1+{n^3+4n^2+56n-96+(n^2-24)\sqrt
{n^2-24n+48}\over 4(n^3+4n^2-24n+144)}\epsilon +O(\epsilon ^2).
\eqno (4.8)$$
\par
Chirality defined here  is a quantity {\it quadratic\/} in spin 
variables. At the standard $O(n)$ Wilson-Fisher fixed point,
there is only {\it one\/} crossover exponent at quartic order in the
spins, namely, the standard anisotropy-crossover exponent. At the
$O(n)$ chiral fixed point, 
as a reflection of richer underlying symmetry,
there generally exist {\it four\/} different crossover
exponents even at the quadratic level, which physically 
represent  {\it   chirality, wavevector-dependent
anisotropy, 
uniform anisotropy   
and wavevector-dependent energy\/} perturbations [23].
Among them, the chiral-crossover exponent 
$\phi _\kappa $ is the largest.
In the particular case of {\it XY\/}  ($n=2$) spins,
the discrete symmetry of the LGW Hamiltonian discussed in \S 3(b)
(the symmetry iii) mixes the two otherwise independent
composite operators,  uniform anisotropy and 
wavevector-dependent energy, and reduces
this number from four  to {\it three \/} [66].
\bigskip\par
\noindent
{\it Effects of commensurability}
\par
\medskip
Under certain circumstances, the LGW Hamiltonian (2.4) could have
terms with a lower symmetry. An example may be seen in the 
$90^\circ $ spiral in helimagnets, where the turn angle 
is just equal to $90^\circ $. In such a case, as 
first noticed by Garel and Pheuty [5], the LGW Hamiltonian has 
an additional
quartic term of the form,
$$w(\vec a^4+\vec b^4). \eqno (4.9)$$
Garel and Pheuty studied the relevance of this quartic term 
by  $\epsilon =4-d$ expansion, and concluded that
this term was relevant in the physical case ($d=3$, $n=2$)  and 
changed the nature of the helical transition from continuous
to first order [5]. In contrast, the present author argued that
this term was irrelevant in the ($d=3$, $n=2$) helical transition
and  even the $90^\circ $ spiral exhibited a
continuous transition of $n=2$ chiral universality [45].
The difference comes from the fact that the fixed points
identified 
by those authors were in fact different: The fixed point 
invoked by Garel and Pheuty  was the Bak and Mukamel
fixed point [4]
while the one invoked
by the present author was the chiral fixed point [23]. 
\bigskip\bigskip
\noindent
(b) $1/n$ expansion
\par
\medskip
In the many-component limit $n\rightarrow \infty $, the
LGW Hamiltonian (2.4) can be solved exactly
for arbitrary dimensionality $d$.
In the noncollinear case $v>0$, on which we shall concentrate in
this subsection, one has a continuous transition
characterized by the standard spherical-model exponents,
$\alpha =(d-4)/(d-2),\ \beta=1/2,\ \gamma=2\nu =2/(d-2)$ for
$2<d<4$ [23].
(In the sinusoidal case $v<0$, the $n\rightarrow \infty $ behavior is
more complex: See Ref.67 for details.)
Thus, in the noncollinear case, 
one can make the standard $1/n$ expansion 
from the spherical model
based on the LGW Hamiltonian (2.4). In the $1/n$ expansion, the
transition is always continuous for $2<d<4$:  First-order
transition found in the $\epsilon =4-d$ expansion for
$n<n_{{\rm I}}(d)$ does not arise.
Various exponents to leading order in $1/n$ were calculated 
as [23]
$$\gamma ={2\over d-2}\{1-9{S_d\over n}\}+O({1\over n^2}),
\eqno(4.10)$$
$$\nu ={1\over d-2}\{1-12{d-1\over d}{S_d\over n}\}+O({1\over n^2}),
\eqno(4.11)$$
{\it etc.\/}, where $S_d$ is defined by
$$S_d=\sin \{\pi (d-2)/2\}\Gamma (d-1)/[2\pi \{\Gamma (d/2)\}^2].
\eqno(4.12).$$
For $n\rightarrow \infty $ and $\epsilon \rightarrow 0$, 
these $1/n$-expansion 
results match the $\epsilon $-expansion 
results {\it obtained at the chiral fixed point\/}.
On comparison with the results 
for the standard $O(n)$ Heisenberg exponents, one sees 
that  both $\gamma $ and $\nu $ of the 
noncollinear transition are smaller than those of the collinear
transition, a tendency consistent with 
the $\epsilon =4-d$ expansion results.
\par
The chiral crossover exponent $\phi _\kappa $ was  calculated 
as [23]
$$\phi _\kappa  =
{1\over d-2}\{1-12{d-1\over d}{S_d\over n}\}+O({1\over n^2}).
\eqno(4.13)$$
Comparison with the  expression for $\gamma $ 
shows that the chiral crossover exponent  exceeds
the susceptibility exponent $\gamma $, although
it is smaller than the gap exponent $\Delta $. Note that the
same inequality  is also satisfied 
within the $\epsilon =4-d$ expansion 
at
the chiral fixed point.
This inequality is somewhat unusual since
in usual cases  crossover exponents have satisfied the inequality
$\phi \leq \gamma $. The 
complete spectrum of crossover exponents at the
quadratic level of spins was  given
in Ref.23.
\par
A modified version of the $1/n$ expansion called self-consistent
screening approximation, in which the standard $1/n$ expansion 
is extended to smaller values of $n$ in a self-consistent manner,
was made by Jolicoeur [68].  A
continuous transition characterized by the exponents different from
the standard $O(n)$ exponents was also found, supporting the
existence of  chiral universality class. 
\par\bigskip
\noindent
(c) What happens in physically relevant cases $d=3$ and $n=2$ or 3?
\par\medskip
Now, in view of the 
$\epsilon =4-d$ and $1/n$ expansion results
presented in the previous subsections, I wish to consider the
physically relevant 
situation, 
$d=3$ and $n=2$ or 3. 
Implications from  the $1/n$ expansion or its extended version 
is simple: A new type of continuous transition 
characterized by the exponents different from those of the standard
$O(n)$ exponents is suggested [23,68].  
Implications from the $\epsilon =4-d$ 
expansion is more subtle, which was summarized in Fig.8.
In the regime $n>n_{{\rm I}}(d)=21.8-23.4\epsilon ^2+7.1\epsilon ^3$,
there occurs a continuous  transition  governed by 
a new chiral fixed point.
By contrast, for $n_{\rm I}(d)>n>n_{{\rm III}}(d)=2-\epsilon +
1.3\epsilon ^2$, there is no stable fixed point in the noncollinear
region and the transition is expected to be first order.
Finally, for $n<n_{{\rm III}}(d)$, the transition is governed by the
standard $O(2n)$ Heisenberg fixed point. 

At $d=3$ and $n=2$ or 3,
this last possibility, {\it i.e.\/}, the noncollinear transition
governed by the
$O(2n)$ Heisenberg fixed point, might be excluded, 
partly because  all RG calculations agree in that
the borderline value $n_{{\rm III}}(d)$ lies below $n=2$ [23,61,63], 
but also
because such $O(2n)$ Heisenberg behavior 
has not been 
seen in extensive Monte Carlo simulations performed on the
stacked-triangular antiferromagnets [21,33-37] (Monte Carlo
results will be reviewed
in the next section).
\bigskip\par
\noindent
{\it Continuous vs. first order}
\par\medskip

Then, the remaining question is whether the transition is  continuous
governed by the chiral fixed point, or it is first order.
Of course, one can also imagine the borderline situation, 
{\it i.e.\/},
the ``tricritical'' case. Possible
tricritical behavior will be discussed separately
in the next subsection.
The above question is equivalent to determining the fate of the
boundary, $n_{\rm I}(d)$, at $d=3$.
As mentioned, previous authors exposed different opinions about
this point. In Ref.23 the present author conjectured that 
$n_{{\rm I}}(3)\leq 2$ 
by invoking the Monte Carlo results.  
Antonenko, Sokolov and Varnashev claimed
that the transition was first order based on their Borel-Pad\'e
estimate, $n_{\rm I}(3)\sim 3.39$, which was slightly larger than the
physical value, $n=3$ [61].
The series  for $n_{\rm I}(d)$ 
used in the resummation procedure, however,  has only three terms, and
as we have seen in the previous subsection
in the {\it XY\/} sinusoidal case,
it is sometimes 
dangerous to draw a definite
conclusion based on such a short series. 
At  present, it would be fair to say that no definite
conclusion could be drawn 
from the $\epsilon $ expansion.
Naively, on may feel 
that the borderline value of $n_{\rm I}$ at the
lowest order, $n_{\rm I}(0)\simeq 21.8$, 
is large enough as compared with the
physical values $n=2$ or 3 so that one may safely conclude 
that the transition in real systems is first order.
However, the coefficient of the first correction term, 23.4, is also
large, which sets the scale
of the numerics in this problem. For example,
the difference between the Borel-Pad\'e estimate of Ref.61 
$n_{\rm I}\simeq 3.3$ and the physical value $n=3$ 
is so small
compared with this scale that one can hardly hope to get a
reliable answer, 
especially
without the knowledge of the asymptotic behavior of the series.

In this connection, it might be instructive to point out that 
an apparently similar situation exists in the phase transition of
lattice superconductors ($U(1)$ lattice gauge model)
with $n$-component order parameter, where the real system corresponds
to $n=2$ [68]. 
A RG $\epsilon =4-d$ expansion calculation applied to this
model yielded a stable fixed point only for very large $n>183$, below
which there was no stable fixed point [69]. 
Since this border value of $n\sim 183$ was so large compared with the
physical value $n=2$, it was initially concluded that the normal-super
transition of charged superconductors should be  first
order [69]. However, it is now well established  through
the duality analysis and  Monte Carlo simulation that the $n=2$ 
superconductor in fact shows  a continuous transition of the 
inverted-{\it XY\/} type [70,71]. 
So, the low-order $\epsilon =4-d$ 
expansion clearly gives a wrong answer in this case. 
By contrast, $1/n$ expansion and its modified version (self-consistent
screening approximation) correctly yielded 
a continuous transition [69,72].

Presumably, the only way in which one could get more or less
reliable answer from the RG loop expansion is to obtain
large-order behavior of the series (large-order perturbation),  
possibly with a few more terms
in the expansion [73].
We leave such a calculation applied to the noncollinear transition
to future studies.
\par
It might also be important to point out here that,
even when a stable fixed point exists as in the regime (I), 
a first order transition is still possible depending on the
microscopic parameters of the system. This is simply due to the
fact that  even in the type of the RG flow diagram in 
Fig.7a the  flow could show a runaway 
only if the initial point 
representing a particular microscopic system is located 
{\it outside\/} the
domain of attraction of the stable fixed point.
This means that, even if one has a few noncollinear systems 
exhibiting a first-order transition, it does not 
necessarily exclude the possibility of a group of other noncollinear
magnets showing a continuous transition.
The difference between these two types of systems
is not of symmetry
origin, but arises simply from the difference in certain 
nonuniversal parameters.
\par
One might then hope to
get
information about the location of the initial point of the RG
flows in the 
parameter space, by mapping the original microscopic spin
Hamiltonian into the LGW form. Of course, 
there usually remains some ambiguities in the procedure
because such a mapping also generates
higher-order irrelevant terms in the LGW Hamiltonian (various
terms higher than sixth
order in $\vec a$ and $\vec b$),  which modifies the initial values of
the quartic terms $u$ and $v$ somewhat 
through a few initial RG iterations.
Anyway, such a mapping performed in Ref.23
shows that  in both cases 
of stacked-triangular
antiferromagnets and helimagnets one has $v_0/u_0=4/3$, where $u_0$
and $v_0$ are the initial values of quartic coupling constants. 
In the situation where the chiral fixed point exists at all, 
this point is
likely to lie {\it inside\/} the domain of the  fixed point.
Indeed, the ratio $v/u$ at the chiral fixed point in the
borderline case $n=n_{{\rm I}}$
is estimated by the $\epsilon =4-d$ expansion as
$v_0/u_0=3.11+O(\epsilon )$.

By contrast,  there are several models with the same chiral symmetry
whose initial point of the RG flow lies outside 
the domain of attraction of the
chiral fixed point. An example may be
the matrix $O(2)$ model
describing the $n=2$ noncollinear magnets, in which 
the noncollinear structure is 
completely rigid. In this model, the above 
mapping yields the initial point at
$v_0/u_0=4$ [74], 
which is expected to lie {\it outside \/}
the domain of attraction of the
chiral fixed point. Here, recall
that the line $v/u=4$ corresponds to the
stability boundary in the mean-field approximation as shown
in \S 4(a), and  is
likely to lie outside the domain of attraction of any stable
fixed point. In fact, a  first-order transition was observed
for the matrix $O(2)$ model in $d=3$ dimensions 
by Monte Carlo simulation [75], 
consistent with the above
argument. In the $O(3)_L\times O(2)_R$ matrix model
representing the completely rigid $n=3$ noncollinear magnets,
the above mapping yields 
$v_0/u_0=3$ [74]. For this matrix model in $d=3$ dimensions,
Kunz and Zumbach observed by Monte Carlo simulation
a continuous transition with unusual critical exponent 
$\nu \sim 0.48$ [75].
For the stacked-triangular Heisenberg antiferromagnet, 
Dobry and Diep  observed by Monte Carlo 
simulation 
that, if one  
stiffened the noncollinear 120$^\circ $ structure by
adjusting some of the exchange constants, the nature of the
transition apparently
changed significantly [76]. This observation might also be 
understandable 
within the above picture,  
if one regards the initial point of the RG flow
moving in the parameter space
toward a runaway
region as  the noncollinear 
120$^\circ $ spin structure is stiffened.
\par
Since there appears to be a possibility that $n_{\rm I}(d=3)$ 
lies close to the
physical values $n=2$ or 3, it may be interesting to examine
what happens if  $n_{\rm I}(3)$ is only very slightly 
larger than the physical value of $n$.
In this case, although 
there is no stable fixed point in
the strict sense (chiral fixed point becomes complex-valued
in this regime),   
RG flows behave as if there were a stable fixed
point for a long period of iterations.
Thus, as
illustrated in Fig.9, a ``shadow'' of
the chiral fixed point attracts the RG flows up to a certain scale, 
but
eventually, the flow escapes away from such a ``pseudo-fixed point''
through a narrow channel in the parameter space and shows a runaway
signaling a first-order transition. Physically,
this means that the system exhibits a rather well-defined critical
behavior for a wide range of
temperature 
governed by
the complex-valued chiral fixed point,  but eventually, 
the deviation from such 
critical behavior sets in for sufficiently small $t$, 
and the system exhibits
a weak first-order transition.
This scenario is perhaps close to the `almost continuous
transition' scenario proposed by Zumbach [77,78]. 
It was suggested there within the
local potential approximation of RG  that the transition of
$n=3$ noncollinear magnets might be almost continuous with 
well-defined pseudocritical exponents.
\par
\bigskip
\noindent
{\it Possible tricritical behaviors}
\par\medskip
A few authors have suggested that the $d=3$ noncollinear transition
might be tricritical. More specifically, 
{\it mean-field\/} tricritical behavior 
was invoked in those works [31,32,36]. 
It should be noticed, however,  that the
tricriticality in general is not
necessarily mean-field tricritical, particular when the 
LGW Hamiltonian has more than one quartic coupling as in our 
model [79]. In this subsection, I will examine the possible tricritical
behaviors in the noncollinear transition 
based on the LGW Hamiltonian (2.4)
and the $\epsilon =4-d$ expansion picture.
Since  the word `tricritical' has sometimes 
been used in the literature
in a rather wide or vague sense, I will try to be unambiguous here
what is meant by tricriticality.
The two different `tricritical' cases will be discussed. 
\par
The standard tricritical situation
is concerned with a separatrix of
the RG flows which divides the two regions of the parameter space,
one associated with a continuous transition 
and the other with
a first-order transition. In the case where the
chiral fixed point is stable, this separatrix is the line connecting
the Gaussian fixed point $G$ and the
antichiral fixed point $C_-$, the latter being
the tricritical fixed point: See Fig.7a. 
By its definition, the tricritical fixed point
has
one more relevant operator in addition to the temperature
and the ordering field. Thus, if the initial Hamiltonian happens to
lie at a point on this separatrix,  RG flow is attracted to the
tricritical fixed point $C_-$ and the system exhibits a 
tricritical behavior governed by the antichiral fixed point $C_-$.
In order to reach this tricritical fixed point, 
one has to tune one symmetry-unrelated microscopic parameter so that
the initial point is just on the separatrix. 
Since the tricritical fixed point here is not the
Gaussian fixed point $G$, 
but the  nontrivial antichiral fixed point $C_-$,
the associated tricritical exponents are {\it not\/}
of mean-field tricritical. As usual, a change
in certain nonuniversal parameter of the system 
would induce either a first-order
transition or a continuous transition governed by the stable chiral
fixed point $C_+$.
\par
The second `tricritical' case is concerned with the situation
where the physical value of $n$ is just at the
borderline value $n=n_I(d)$ between the regimes of continuous and
first-order transitions.
In this case,
the RG flow diagram becomes as given in Fig.10, where
the two fixed points $C_+$ and $C_-$ coalesce
at a point in the $(u,v)$ plane. 
As can be seen in Fig.10, the resulting fixed point, which
is again a highly nontrivial one, has 
a finite domain of attraction in the $(u,v)$ plane and attracts
many microscopic Hamiltonians, in contrast to the tricritical
fixed point discussed above.
Therefore, except for the degenerate nature of the fixed point,
the situation is essentailly the same as in the case of
$n>n_{\rm I}(d)$, in the sense that 
novel critical behavior is expected
for a variety of microscopic systems.
\par
Note that, in either case discussed above, 
the tricritical behavior is highly
nontrivial, {\it not of  mean-field tricritical\/}. 
This is simply
due to the fact that the  tricritical fixed point 
is a nontrivial one
reflecting the existence of more than one
quartic coupling constant in the LGW Hamiltonian.
Of course, the Gaussian fixed point responsible for the mean-field
tricritical behavior always exists at the
origin, but to reach this fixed point,
one has to tune more than one symmetry unrelated microscopic
parameters, and
the occurrence of such 
mean-field tricritical transition is highly unlikely [79].
\par
\bigskip
\noindent
(e) $\epsilon =d-2$ expansion
\par
\medskip
In this subsection, I will review an alternative RG approach,
an expansion from the lower critical dimension $d_<=2$.
The application of this method to the noncollinear transition was first
made by Azaria, Delamotte and Jolicoeur 
for the Heisenberg spins ($n=3$) [31].
Extension to general $n$-component spins was made by 
Azaria, Delamotte, Delduc  and Jolicoeur [32], 
and by the present author [80].
\par
\bigskip
\noindent
{\it Nonlinear sigma model\/}
\par
\medskip
In contrast to the $\epsilon =4-d$ expansion, 
the $\epsilon =d-2$ expansion is based on the nonlinear sigma model
which is written in terms of spin variables of
fixed length. In case of noncollinear 
magnets with $n$-component spins, 
this may be written in terms of
two mutually orthogonal $n$-component vector
fields $\vec a$ and $\vec b$ as
$${\cal H}={1\over 2T}[(\bigtriangledown_\mu \vec a)^2+
(\bigtriangledown_\mu \vec b)^2+r\sum_{1\leq i<j\leq n}
\{ \bigtriangledown_\mu (a_ib_j-a_jb_i)\} ^2], \eqno(4.14a)$$
with the constraints
$$\mid \vec a({\bf r})\mid =\mid \vec b({\bf r})\mid =1,\ \ \ \  
\vec a({\bf r})\cdot \vec b({\bf r})=0,\eqno(4.14b)$$
where $T$ is a temperature and $r$ is a coupling-constant ratio.
One can easily check that the above Hamiltonian satisfies the
same 
$O(n)\times O(2)$ symmetry as in the LGW Hamiltonian (2.4).
Unlike the case of eq.(2.4),  
the noncollinear structure, {\it i.e.\/}, an orthogonal
frame spanned by the two vectors $\vec a$ and $\vec b$, is
completely rigid here. 
It is
not necessarily obvious whether this idealization does not
change  essential physics in $d=3$ dimensions
(Recall our discussion 
concerning
the stiffness of the noncollinear structure
in the previous subsection based on the LGW Hamiltonian).
\par
\bigskip
\noindent
{\it Fixed points and exponents\/}
\par
\medskip
The standard $\epsilon =d-2$ expansion applied to the
Hamiltonian (4.14) yields a stable fixed point characterized by
the exponents [32,80]
$$\nu =\epsilon -{1\over 2}{6n^3-27n^2+32n-12\over (n-2)^3(2n-3)}
\epsilon ^2+O(\epsilon ^3),\eqno (4.15)$$
$$\eta ={3n^2-10n+9\over 2(n-2)^3}\epsilon +O(\epsilon ^2).
\eqno (4.16)$$
This fixed point is stable for any $n>2$ and $d>2$. In the limit
$n\rightarrow 2$, the fixed-point temperature tends to infinity and
the $\epsilon =d-2$ expansion becomes meaningless.
Azaria {\it et al\/} observed that, in the particular case 
of Heisenberg 
spins ($n=3$),  
the obtained fixed point was nothing but
the standard $O(4)$ Wilson-Fisher 
fixed point [31]. Note that this $O(4)$ fixed point is different in nature
from the $O(4)$-like fixed point obtained by Bak and Mukamel in the
$\epsilon =4-d$ expansion analysis of the {\it XY\/}  ($n=2$) 
noncollinear magnets:
The former fixed point has no counterpart in the $\epsilon =4-d$
expansion [4]. By contrast, for $n>3$, the  fixed point 
obtained by the $\epsilon =d-2$ expansion is a new one,
{\it not\/} the standard Wilson-Fisher fixed point. Indeed,
for large enough
$n$, various exponents reduce to those obtained by the 
$1/n$ expansion based on the LGW Hamiltonian [23],
naturally fitting into the chiral-fixed point picture obtained
by the $\epsilon =4-d$ and $1/n$ expansions.
\par
As mentioned, in the Heisenberg ($n=3$) case,  $\epsilon =d-2$ 
expansion predicts that
the symmetry is dynamically restored,
yielding the standard $O(4)$ critical behavior
which has never been seen in the $\epsilon =4-d$ expansion.
Based on this observation,  Azaria {\it et al\/} claimed
that the ($n=3$, $d=3$) noncollinear transition should be
of  standard $O(4)$ universality [31,32].
They further speculated that the transition could also be first
order or mean-field tricritical, depending on the microscopic
parameters of the system.  (Note, however, 
that the $\epsilon =d-2$ expansion
itself yielded neither  first-order nor 
mean-field tricritical behavior.) So, in the 
`nonuniversality' scenario of Ref.31, the noncollinear transition 
of Heisenberg systems is either $O(4)$, mean-field tricritical, or
first order. 
\par
\bigskip
\noindent
{\it Discussion\/}
\par
\medskip
In fact, as will be
shown in the next section, recent extensive Monte Carlo
simulations on the stacked-triangular Heisenberg
antiferromagnets now rule out
the $O(4)$-like critical behavior [31,33,35,37]. 
Thus, doubt has been cast  by several authors to the validity of
the $\epsilon =d-2$ method applied to this problem.
In the Heisenberg case ($n=3$), 
a different interpretation of the $O(4)$ behavior obtained by the
$\epsilon =d-2$ expansion
had already been exposed in Ref.80:
It was argued there that the $O(4)$ fixed point
for $n=3$
was  spurious, arising from the 
incapability of the  method to deal 
with the crucially 
important nonperturbative effects associated with the 
vortex degrees of
freedom, which reflects
the nontrivial topological structure
of the order-parameter space, $\Pi _1(V=SO(3))=Z_2$.
Essentially the same criticism was also made by Kunz and Zumbach,
and by Zumbach 
in Refs.75 and 81.
\par
By analyzing the properties of another generalization
of the $n=3$ model, $O(n)\times O(n-1)$ nonlinear sigma model,
David and Jolicoeur proposed a scenario in which  Azaria's
$O(4)$ fixed point with enlarged symmetry played no role due to the
appearance of a first-order line in the phase diagram [82].
(Note that the `principal chiral fixed point' quoted by these authors
corresponds to the $O(4)$ fixed point 
with enlarged symmetry, {\it not\/}
the chiral fixed point in the present article.)
On the other hand, based on their Monte Carlo study of 
a modified stacked-triangualr Heisenberg antiferromagnet 
in which
the interaction is modified to yield the rigid $120^\circ $ structure,
Dobry and Diep 
suggested that the nonlinear sigma model used by Azaria 
{\it et al\/} itself might already be inappropriate to model
the original stacked-triangular Heisenberg antiferromagnet [76].
\par
While the above criticisms apply specifically 
to the $n=3$ noncollinear magnets, it should also be mentioned that
there has been a  controversy 
concerning the validity of
the $\epsilon =d-2$ expansion method 
even in the simplest case of simple $O(n)$ ferromagnets 
[83].
Anyway, it now appears clear  in the present problem that 
the $\epsilon =d-2$ expansion method is problematic, 
at least in the case of $n=3$.  Special care 
has to be taken in applying this method
to the system with nontrivial internal structure in its
order-parameter space like the 
noncollinear magnets.
\par
\bigskip
\noindent
(f) Further generalization of noncollinear transitions
\par
\medskip
So far, we have limited our discussion 
to the magnets with the noncollinear but {\it coplanar\/} spin order.
On the other hand, in some cases, {\it noncoplanar\/} 
spin orderings that 
are  three-dimensional in spin space could appear.
Example is a triple-$\vec Q$ ordering as 
illustrated in Fig.11. One can further generalize the situation
to $m$-dimensional spin order in isotropic $n$-spin space with
$m\leq n$. The $m=1$ case represents the collinear spin 
order, while the $m=2$ case represents the noncollinear but
coplanar spin order  discussed so far. Then, one can
naturally imagine  the possible existence of  
hyperuniversality series
characterized by two integers ($m,n$).

Theoretical analysis of such noncoplanar criticality
was first made in 1990 by the present author
based on a symmetry argument, RG $\epsilon =4-d$ and
$1/n$ expansions [84]. 
An appropriate LGW Hamiltonian with
the $O(m)\times O(n)$ symmetry is given by
$${\cal H}_{{\rm LGW}}
={1\over 2}\sum _\alpha (\nabla \vec \phi _\alpha )^2+
{1\over 2}r\sum _\alpha \vec \phi _\alpha ^2+
{1\over 4!}u(\sum _\alpha  \vec \phi _\alpha ^2)^2 + 
{1\over 4!}v\sum _{<\alpha \beta>} \{(\vec \phi _\alpha 
\cdot \vec \phi _\beta )^2-\vec \phi _\alpha ^2 \vec \phi 
_\beta ^2\}, \eqno (4.17)
$$
where {$\vec \phi_\alpha $} ($1\leq \alpha \leq m$) are $m$ sets of
$n$-component vectors. The condition 
$$0<v<{2m\over m-1}u,\eqno (4.18)$$
is required by the noncoplanarity of the ordering and the boundedness
of  free energy. The $\epsilon =4-d$ expansion applied to (4.17)
yields a generalized chiral fixed point in the noncollinear
region $v>0$, which is stable for [84]
$$n>n_{{\rm I}}(d)=5m+2+2\sqrt {6(m+2)(m-1)}-\{5m+2+
{25m^2+22m-32\over 2\sqrt {6(m+2)(m-1)}}\}\epsilon
+O(\epsilon ^2).\eqno (4.19)$$
For the coplanar ($m=2$) case, 
this reduces to the previous result (4.3), while
in the noncoplanar ($m=3$) case, this gives
$$n>n_c(d)=32.5-33.7\epsilon +O(\epsilon ^2).\eqno (4.20)$$
Again, it is not easy to tell from this
expression
whether the noncoplanar ($m=3$) 
chiral fixed point remains
stable in the physical case,  $d=3$ and $n=3$.

The exponents $\gamma $ and $\nu $ 
at this generalized chiral fixed point
were calculated  as [84]
$$\gamma \approx 2\nu =1+{1\over 4}B_{mn}(C_{mn}+D_{mn}\sqrt{R_{mn}})
\epsilon +O(\epsilon ^2),$$
$$B_{mn}^{-1}=(mn+8)(m+n-8)^2+24(m-1)(n-1)(m+n-2),$$
$$C_{mn}=mn(m+n)^2+8mn(m+n)-22(m+n)^2+
88mn-32(m+n)+152,$$
$$D_{mn}=mn(m+n)-10(m+n)+4mn-4,$$
$$R_{mn}=(m+n-8)^2-12(m-1)(n-1). \eqno (4.21)$$
The $1/n$ expansion applied to (4.17) yields a continuous transition
characterized by the  exponents [84],
$$\gamma ={2\over d-2}\{1-3(m+1){S_d\over n}\}+O({1\over n^2}),
\eqno (4.22)$$
$$\nu ={1\over d-2}\{1-4(m+1){d-1\over d}{S_d\over n}\}
+O({1\over n^2}),\eqno (4.23)
$$
where $S_d$ was defined by (4.12).
Further details including the expression of the
chiral crossover exponent
were given in Ref.84. (A part of the 
$\epsilon =4-d$
expansion results at the lowest order 
was also reported in Ref.85, in apparent ignorance of Ref.84.)
Anyway, if this generalized chiral fixed point remains stable
in $d=3$,
the associated critical behavior is most probably novel.
Thus,  possible
existence of a hyperseries of universality classes characterized
by two integers $m$ and $n$ was proposed in Ref.84, where the
special case $m=1$
corresponds to the standard $O(n)$ Wilson-Fisher
universality and the case $m=2$
corresponds to the standard chiral universality.

One possible example of such noncoplanar criticality was 
studied by
Reimers, Greedan and Bj\"orgvinsson for  pyrochlore 
antiferromagnet FeF$_3$ both by neutron-diffraction
experiment and  by 
Monte Carlo simulation [86]. The reported exponent 
values were quite unusual, $\alpha =0.6(1),\ \beta =0.18(2),\ 
\gamma =1.1(1)$ and $\nu =0.38(2)$, although Mailhot and Plumer
argued that the same data were also not inconsistent with a
first-order transition [87].  
\par
\bigskip
\bigskip
\noindent
\S5 {\bf Monte Carlo simulations}
\par
\medskip
\noindent
(a) Stacked-triangular antiferromagnets
\medskip
In this section, I wish to review the  results of
Monte Carlo simulations on the 3D {\it XY\/} and 
Heisenberg antiferromagnets
on a stacked-triangular lattice. 
Monte Carlo method enables us to study 
the {\it XY\/} and Heisenberg  systems directly
in three dimensions. Thus, if one could control  
finite-size effects and   statistical errors
intrinsic to the method, one could get useful
information which might serve to test various theoretical
proposals.
\par
Partly for simplicity and partly to get a wide
critical regime, most of  extensive Monte Carlo simulations
on  the stacked-triangular
antiferromagnets were performed on the nearest-neighbor
Hamiltonian (2.1) with $J=J'$.
Earlier work by the present author simulated the 
lattices  up to $L=60^3$ both for 
$XY\/$ and Heisenberg models based on the conventional method [21],
while more recent simulations on the {\it XY\/} model
by Plumer and Mailhot [36], by Boubcheur, Loison and Diep 
[34], and those on the Heisenberg model
by Bhattacharya, Billoire, Lacaze and Jolicoeur [33],  
by Mailhot, Plumer and Caill\'e [37], 
and by Loison and Diep [35] 
used the histogram technique, the largest
lattice sizes being $L=33\sim 48$.
As an example,  the temperature and size dependence of the
specific heat calculated in Ref.21
is reproduced in Fig.12.
In the numerical sense,
the results obtained by these independent simulations
agreed with each other except for a small deviation left
in some exponents in the {\it XY\/} case.
All authors observed a continuous transition both for
the {\it XY\/} and Heisenberg cases, except for a recent simulation
by Mailhot and Plumer on a {\it quasi-one-dimensional\/} 
stacked-triangular {\it XY\/} antiferromagnet [88].

The  values
of critical exponents, specific-heat amplitude ratio and 
transition temperature 
reported by these authors
are summarized in Table II and III
for both cases of {\it XY\/} and Heisenberg models, 
and are compared with the corresponding 
values of unfrustrated {\it XY\/} and Heisenberg
ferromagnets, 
of the standard $O(4)$ behavior,
and of  the mean-field 
tricritical behavior. 
One can immediately see that the exponent values
determined by these simulations 
differ significantly from the   unfrustrated {\it XY\/} or
Heisenberg values.
One can also see  that
the reported exponents are incompatible with the 
$O(4)$ exponents  both in the {\it XY\/} and Heisenberg cases,
which were
predicted by Bak and Mukamel in the {\it XY\/} case [4] and
by Azaria {\it et al\/} in the Heisenberg case [31]. Indeed, the
$O(4)$ singularity is weaker than that of the standard {\it XY\/} and 
Heisenberg singularity,
contrary to the observed tendency.
Based on these findings, 
one may now rule out the standard $O(4)$-like critical
behavior in both cases of
{\it XY\/} and Heisenberg magnets.
In the Heisenberg case, the reported exponents are also inconsistent
with the 
mean-field tricritical values suggested by Azaria {\it et al\/} [31], 
and give 
support to the claim
that the $n=3$ noncollinear transition 
is indeed of new $n=3$ chiral universality.  

In the {\it XY\/} case, one sees from the table that
the reported exponent values
are not much different from the mean-field tricritical
values. 
Furthermore, a closer look  reveals that
there remains small difference in
the exponent values reported by three different groups.
All agree concerning the exponent $\beta $ which comes around
0.25.  By contrast, concerning the exponent $\gamma $,
the reported values are scattered 
as $0.99 \pm 0.02$ (Ref.34), $1.13 \pm 0.05$ (Ref.21) and
$1.15\pm 0.05$ (Ref.36).  
The reason of this deviation is not clear.
In fact, the exponent values reported by Plumer and Mailhot
in Ref.34 were very close to the mean-field tricritical values,
and these authors suggested that the transition in the {\it XY\/}
case might indeed be mean-field tricritical. 
In contrast to this,  finite-size scaling analysis in Ref.21
favored the nontrivial exponents, rather than the mean-field
tricritical exponents. 
\par
Meanwhile, larger deviations from the
mean-field values were observed in the chirality exponents
$\beta _\kappa $ and $\gamma _\kappa $ and the specific-heat 
amplitude ration $A^+/A^-$. In the mean-field tricritical case
governed by the Gaussian fixed point,
these values should be $\beta _\kappa =0.5$, $\gamma _\kappa =0.5$ and 
$A^+/A^-=0$, while the Monte Carlo results of Ref.21 yielded
$\beta _\kappa =0.45\pm 0.02$, $\gamma _\kappa =0.77\pm 0.05$ and 
$A^+/A^-=0.36\pm 0.2$ in the {\it XY\/} case, and 
$\beta _\kappa =0.55\pm 0.04$, $\gamma _\kappa =0.72\pm 0.08$ and 
$A^+/A^-=0.54\pm 0.2$ in the Heisenberg case. 
These nontrivial values of the chirality exponents and the
specific-heat amplitude ratio appear to be hard 
to explain from the
mean-field tricritical scenario. The observed chirality exponents 
satisfy the scaling relation (4.7) within the error bars. 
\par
In Ref.34, Plumer and Mailhot suggested a possibility that the 
chirality and the spin are decoupled and order at slightly different
temperatures, $T_c\neq T_c^{(\kappa )}$, and/or
with mutually different correlation-length exponents, $\nu \neq
\nu _\kappa $. 
From the standard theory of critical phenomena, however,
this is a rather unlikely situation in the present 3D problem
due to the following reason.
If the chirality were decoupled from the spin and exhibited an
independent transition, the criticality associated with this chirality
transition is expected to be of 3D Ising universality, which then
should give $\beta _\kappa \sim 0.324$, $\gamma _\kappa \sim 1.239$
and $\nu _\kappa \sim 0.629$ {\it etc\/}. However, this clearly 
contradicts the Monte Carlo results.
Even if the criticality of the
decoupled chirality transition were to differ from the standard
Ising one due to some unknown reason, the chiral susceptibility
exponent $\gamma _\kappa $ in such a case
should definitely  be larger than unity,
which again seems hard to reconcile with 
the Monte Carlo results $\gamma _\kappa 
=0.77\pm 0.05$ [21] or $\gamma _\kappa =0.90\pm 0.09$ [34].
Rather, the Monte Carlo observation that $T_c\sim T_c^{(\kappa )}$ 
and  $\nu \sim \nu _\kappa $, together with the non-Ising 
values of the chirality exponents is a  clear 
indication that the spin
and the chirality are {\it not\/} decoupled and the chirality 
behaves as
a composite operator of the order parameter, the spin. 
In fact, this  is just a scenario suggested from 
the RG analysis in \S 4(b) [23]. Note that, in such a situation, 
the  chirality
exponents are generally non-Ising and the chiral susceptibility
exponent $\gamma _\kappa $
could be less than unity, in accord with
the Monte Carlo results.
As long as 
the spin and the chirality are not decoupled at the
transition, the observed nontrivial values of the chirality exponents
are unambiguous indications that the transition here is  
{\it not} of mean-field
tricritical.
\par

Monte Carlo simulation
is performed for finite systems (in the present case, $L\leq 60^3$),
and one cannot completely rule out the possibility that a sign 
of  first-order transition
eventually develops for still larger lattices. 
Mailhot and Plumer recently performed a histogram Monte Carlo
simulation of a quasi-one-dimensional stacked-triangular {\it XY\/}
antiferromagnet in which the interplane interaction is much stronger
than the intraplane interaction ($J'=10J$) for  lattice
sizes up to $L=33^3$, and claimed that the
transition  was weakly first order [88]. More specifically,
these authors estimated the transition temperature by  two 
different
methods which gave somewhat  different estimates of $T_c$
(about 0.5\% difference).
If a higher estimate of $T_c$ was employed in the fit,
finite-size scaling of the data was 
suggestive of a first-order transition, while
if a lower estimate of $T_c$ was employed, it was suggestive of
a continuous transition with the exponents close to the previous 
works [34]. In view of the rather large uncertainty 
in their estimate of
$T_c$ as well as  high sensitivity of the results
on the assumed $T_c$ value, and also of the fact that they never observed a
double-peak structure in the energy histogram characteristic of a
first-order transition [88], 
the claimed first-order nature of the
transition appears not necessarily conclusive. One should also be
careful that, in highly anisotropic systems like the one
studied in Ref.88,
there generally occurs a dimensional crossover which
might complicate the data analysis particularly
when the system size is not large
enough.
\par
\bigskip
\noindent
(b) helimagnets
\medskip
While the  stacked-triangular 
antiferromagnets are the best studied model,
there are a few Monte Carlo works on 3D helimagnets (spiral magnets).
Diep  simulated a helimagnetic model with
the competing nearest- and next-nearest-neighbor 
antiferromagnetic interactions on a body-centered-tetragonal 
lattice under  periodic boundary conditions [89]. 
In the case of Heisenberg spins, Diep observed a continuous transition
characterized by the exponents 
$\alpha =0.32\pm 0.03$ and $\nu =0.57\pm 0.02$, which were not
far from the $n=3$ chiral values obtained for the stacked-triangular
Heisenberg antiferromagnet. 
In the case of {\it XY\/} spins, he
observed either two successive continuous 
transitions or a first-order transition, 
depending on the microscopic
parameters of the model.
 
One potential problem exists, however, 
in the simulation of helimagnets of this type. Namely,
unlike the $120^\circ$ spin structure in the triangular
antiferromagnets, the pitch of  magnetic spiral is 
generally {\it temperature
dependent\/} and is {\it incommensurate\/} 
with the underlying lattice.
Therefore,  imposed periodic boundary conditions, even if
they are chosen to accommodate the ground-state spin configuration 
without mismatch, generally
causes a mismatch 
around $T_c$ causing 
an artificial ``stress'' on the helical spin structure. 
This could give a significant effect on the nature of  
phase transition [90], particularly when
the lattice size is not large enough compared with the spiral pitch.
\par
\bigskip
\noindent
(c) Matrix models
\medskip
Finally, several matrix models expected to  
model the noncollinear magnets were also studied by
Monte Carlo simulations.  Hamiltonian of these matrix models
may be given by
$${\cal H}=-J\sum_{<ij>}{\rm Tr}(O_i^TO_j), \eqno (5.1)$$
where $O_i$ is a matrix variable at the $i-$th site of a
simple cubic lattice 
and $J>0$ is the ferromagnetic nearest-neighbor coupling. 
Relevant to our present study is
the matrix $O(2)$ model representing the noncollinear
{\it XY\/} magnets, where
the matrix variable $O_i$ is a $2\times 2$ orthogonal matrix, 
and the
matrix  $O(3)_L\times O(2)_R$ model
representing the noncollinear Heisenberg 
magnets, where $O_i$ is a $3\times 2$ matrix 
written in terms of two orthogonal
unit three-vectors, $\vec a$ and $\vec b$, as ($\vec a,\ \vec b$).

In the $O(2)$ case, 
the model has an $O(2)_L\times O(2)_R$ symmetry and
is also equivalent to
the coupled Ising-{\it XY\/} model of the form
$${\cal H}=-J\sum _{<ij>}(1+\sigma _i\sigma _j)\cos (\theta _i-
\theta _j),\eqno (5.2)$$
where $\sigma _i=\pm 1$ is an Ising variable and $\theta _i
=[0,2\pi )$ is
an angle variable of the {\it XY\/} spin.

As mentioned, these matrix models represent  completely
rigid noncollinear spin structures. Analysis in \S 4 suggests that
these models, particularly the matrix
$O(2)$ model, are likely to 
exhibit 
a first-order
transition since the initial point of the associated RG flow 
might be located 
in the runaway region in the parameter space. 
Indeed, Monte Carlo simulations by Kunz and Zumbach [75], and by
Dobry and Diep [76], 
on these and related models revealed that
the 3D matrix models 
exhibited a first-order transition, or the behavior close to it. 

As pointed out by Zumbach [78],  
the matrix O(2) model shows an interesting transition
behavior {\it even at
the mean-field level\/},  significantly different  
from that of  stacked-triangular antiferromagnets with the non-rigid
noncollinear spin structures: 
It exhibits a mean-field tricritical transition with the exponents
$\alpha =1/2$, $\beta =1/4$\ and $\gamma =1$,
which should be
contrasted to ordinary mean-field exponents 
$\alpha =0$, $\beta =1/2$\ and $\gamma =1$  observed when the
mean-field approximation is applied to the
{\it XY\/} and Heisenberg stacked-triangular antiferromagnets.
By contrast, the $O(3)_L\times O(2)_R$ matrix model
modeling the  rigid Heisenberg noncollinear
magnets  
exhibits an ordinary
mean-field transition  at the mean-field level [78].
These observations suggest that the
nature of the transition of the matrix models or the coupled
Ising-{\it XY\/} model 
may not always be the 
same as those of  original noncollinear magnets with 
the nonrigid 
noncollinear spin structures, even when both share
the same symmetry.
\par
\bigskip\bigskip
\noindent
{\bf \S 6. Experiments}
\par
\bigskip
In this section, we briefly review the
recent experimental results both on (a) stacked-triangular
antiferromagnets and (b) helimagnets (spiral magnets). 
Since some review articles 
with emphasis
on   experimental works are already available [42,44], 
I summarize here some of the main features and 
highlight the points of interest.
\par
\bigskip
\noindent
(a) Stacked-triangular antiferromagnets
\par
\medskip
The best studied material of the stacked-triangular {\it XY\/} 
antiferromagnets is CsMnBr$_3$,
for which  specific-heat measurements (exponent $\alpha $ and 
amplitude ratio $A^+/A^-$) [26,27] and
neutron-scattering measurements (exponents $\beta $, 
$\gamma $ and $\nu $) [24,25,39] 
were made independently by several groups.
The reported values
of the exponents and the specific-heat amplitude ratio are 
summarized in Table IV.   
As an example, the specific-heat data reported in Ref.27 and
the sublattice-magnetization data reported in Ref.24 were 
reproduced in Figs.13 and 14, respectively.
All authors reported a continuous transition. 
In particular, 
high-precision specific-heat measurements 
gave a  stringent upper limit to the possible latent heat,
demonstrating  continuous nature of the transition.
Another example of the well-studied $n=2$ chiral system is
CsNiCl$_3$ under high magnetic fields, for which 
the measured exponents are also 
included
in  Table IV [91-93]. Although CsNiCl$_3$ is 
a weakly Ising-like magnet, under external fields higher than
a certain value $H_m$ corresponding to the multicritical
point, it exhibits a single transition 
directly from the
paramagnetic state to an ``umbrella-type'' noncollinearly-ordered 
state with the
nontrivial chirality. This is caused because  applied
fields generate an effective planar anisotropy perpendicular to
the field, which cancels
and exceeds the intrinsic axial anisotropy.
Overall, 
as can be seen from Table IV, 
the experimental results  support the 
chiral-universality prediction. It should also be
noticed that the measured exponents $\beta $,
$\gamma $ and $\nu $ are not far from the mean-field tricritical
values, although the observed $\nu $ 
marginally favors the nontrivial $n=2$ chiral
value. By contrast, the specific-heat exponent $\alpha $ and the
amplitude ratio $A^+/A^-$ more or less favor the 
chiral-universality values over the mean-field tricritical values.
\par
Relatively well studied Heisenberg-like stacked-triangular
antiferromagnets are
VCl$_2$ [29], VBr$_2$ [28,30], RbNiCl$_3$ [94-96] 
as well as CsNiCl$_3$ in an external field 
corresponding to the multicritical point
($H=H_m$) [91-93]. Note that the former three compounds
are nearly Heisenberg systems,
possessing a weak axial magnetic anisotropy.
The measured values of the exponents and
the specific-heat amplitude ratio are summarized in Table V.
Except for a relatively large deviation observed in the exponent 
$\beta $ and $\gamma $
for VCl$_2$, the results are consistent with 
with $n=3$ chiral values. Since the high-precision specific-heat
measurement for VBr$_2$ yielded  results in good agreement with
the theoretical $n=3$ chiral values, it might be  interesting to
examine the critical properties of VBr$_2$ by neutron
scattering to measure $\beta $, $\gamma $ and $\nu $. 
\par
Other stacked-triangular {\it XY\/}
antiferromagnets studied  are
RbMnBr$_3$ and CsCuCl$_3$. 
Unlike the  compounds quoted above,  
the lattice structures of these  compounds 
around $T_c$  are distorted from the perfect
simple hexagonal lattice.
RbMnBr$_3$ exhibits an incommensurate spin order with
its turn angle equal to 128$^\circ $ [97], presumably due to its
distorted lattice structure [45,46]. 
Concerning the critical properties associated with 
the incommensurate spin
order of RbMnBr$_3$, 
a theoretical argument was given that the critical behavior
would be the same chiral one
as   in
undistorted CsMnBr$_3$ 
if the lattice deformation of
RbMnBr$_3$ is of certain type [45]. 
Indeed, for RbMnBr$_3$, Kato {\it et al\/}
gave $\alpha =0.42\pm 0.16$, $\alpha '=0.22\pm 0.06$ 
and $A^+/A^-=0.30 \pm 0.02$ by  birefringence measurements
[98], 
and $\beta =0.28\pm 0.02$ by  neutron-diffraction measurements [99],
in reasonable agreement with the 
expected $n=2$ chiral values.

By contrast, 
the lattice structure of  CsCuCl$_3$ is distorted such that
the anisotropic Dzyaloshinski-Moriya interaction 
$-\vec D_{ij}\cdot \vec S_i \times \vec S_j$ arises between the
neighboring 
spins along the $c$-axis,  the associated $D$-vector 
pointing to the directions slight off 
the $c$-axis [100].  
Along the $c$-axis, the directions of these $D$-vectors rotate
around the $c$-axis with the period of  six lattice spacings.
If the $D$-vector were precisely parallel with the $c$-axis, 
the spin symmetry would be 
chiral, {\it i.e.\/} $O(2)=Z_2\times SO(2)$, where $Z_2$ concerns
the chiral degeneracy associated with the noncollinear spin
structure {\it in the triangular layer\/}.
However, the  
canting of the $D$-vector from the
$c$-axis
reduces the spin symmetry  from the perfect
chiral one to the lower one, {\it i.e.\/}, 
only $Z_2$ associated with the spin inversion.
Thus, a crossover from the $n=2$
chiral critical behavior 
is expected in its magnetic  transition 
in a close vicinity of $T_c$ [44]. In that sense, 
CsCuCl$_3$ 
is not an ideal material to study the chiral criticality. 

The magnetic phase transition of CsCuCl$_3$  was recently studied
by  neutron diffraction by Mekata {\it et al\/} [101],  by
St\"usser {\it et al\/} [102], and
by specific-heat measurements
by Weber {\it et al\/} [103]. 
Mekata {\it et al\/} obtained $\beta =0.25\pm 0.01$ while
St\"usser {\it et al\/} obtained [102] $\beta =0.23\pm 0.02$, 
which were close to the $n=2$ chiral value and that of CsMnBr$_3$.
By contrast,  Weber {\it et al\/} observed 
in the temperature range 
$10^{-3}<\mid t\mid< 5\times 10^{-2}$ a power-law
scaling behavior in the specific heat
characterized by  
$\alpha =0.35\pm 0.05$ and $A^+/A^-=0.29\pm 0.05$
close to the $n=2$ chiral values, but observed a
deviation from this scaling behavior in a closer vicinity of
$T_c$. This deviation was interpreted by these
authors as a sign of  first-order
transition. It was further 
suggested that this might indicate the failure of
chiral universality.
It should be noticed, however, that due to the reduction of 
spin symmetry caused by the canting of its
$D$-vector from the $c$-axis
CsCuCl$_3$ is not an ideal material
to study the $n=2$ chiral criticality,
and the observed deviation from the $n=2$ chiral critical behavior
might possibly be caused by
the expected crossover effect, 
not being an intrinsic property of an ideal
$n=2$ chiral magnet.  Experimental observation reported in Ref.103 
that external
fields applied 
along the $c$-axis  made the deviation from 
the ideal chiral critical behavior less pronounced 
can naturally be understood from such crossover picture, 
because the $c$-axis field tends to confine the noncollinear spin
structure in a plane orthogonal to the field, thus relatively
weakening the crossover  
due to the canting effect of the $D$-vector.

One should also note that, as 
emphasized in \S 4(c), 
theory leaves enough room for 
the occurrence of a first-order transition even when 
there exists a chiral
universality class.
Hence,  observation of  first-order transition 
in a few noncollinear magnets
is not quite enough to
rule out the possible existence of  chiral universality class
in generic noncollinear transitions.
\par\bigskip
\noindent
(b) helimagnets
\medskip\par
In this subsection, I wish to   review experimental
situation for helimagnets (spiral magnets). 
So far,  experimental studies of the critical properties of these
helimagnets have been limited almost exclusively to
rare-earth helimagnets, Ho, Dy and Tb.
As mentioned in the
Introduction,  experimental
situation for these rare-earth helimagnets has remained  confused.
Different authors
reported considerably different values  for the same exponent of
the same material,
and the reason of this  discrepancy has  not been clear.
Here, I donot intend to give a comprehensive review of various
experimental works, 
but rather highlight several points of
the most severe conflict,  discuss its possible origin
and propose possible ways to disentangle the present
confusion. For  detailed review of the experimental works on
rare-earth helimagnets, we refer the reader to Ref.16.

Let us begin with a  survey of 
the present experimental status. 
Most authors reported that the paramagnetic-helimagnetic transition 
of Ho, Dy and Tb was continuous.
\par\medskip
\noindent
{\it Exponent $\alpha $\/ and  specific-heat 
amplitude ratio $A^+/A^-$}
\par\medskip
Several high-precision
specific-heat measurements have been done on Ho, Dy and Tb.
For Dy, Ledermann and Salamon
reported a crossover from 
the behavior characterized by  $\alpha =-0.02 \pm 0.01$ and
$A^+/A^-=0.48\pm 0.02$ ($10^{-2.3}
<t<10^{-0.5}$) to the
behavior characterized by  $\alpha =0.18 \pm 0.08$ and
$A^+/A^-=0.44\pm 0.04$ ($10^{-3.3}<t<10^{-2.3}$) [7].
Jayasuriya and coworkers
gave $\alpha =0.27\pm 0.02$ and 
$A^+/A^-=1.78\pm 0.45$ 
for Ho [104], $\alpha =0.24\pm 0.02$ and
$A^+/A^-=0.41\pm 0.05$ 
for Dy [105], and $\alpha =0.20\pm 0.03$ and
$A^+/A^-=0.58\pm 0.34$ 
for Tb [106]. Jayasuriya {\it et al\/} 
noticed that the values of $\alpha $
and $A^+/A^-$ 
changed somewhat
depending on the fitting form and the temperature range used in
the fit. 
For Ho, Wang, Belanger and Gaulin
gave $\alpha =0.10\pm 0.02$ and $A^+/A^-=0.51\pm  
0.06$ ($0.002<t<0.1$), or $\alpha =0.22 \pm 0.02$ 
and $A^+/A^-=0.61\pm  0.07$ ($0.002<t<0.1$), 
depending on the particular form of the fitting formula [26]. 
They also
reported that the observed critical behavior could not be well fitted
with a single exponent.
All measurements quoted above 
agreed in that the transition was continuous.
Although there exists considerable scatter among the reported values of
$\alpha $ and $A^+/A^-$, 
a tendency appears clear: The exponent $\alpha $ tends to be 
larger than the standard $O(n)$ values and that
there is a crossover-like behavior which hinders the data to
lie on a single power-law behavior in the temperature range
studied.
\par
There were also several attempts
to extract the specific-heat exponent 
from some other physical quantities such
as  electrical resistivity [107]. Since the validity of such
procedure was  questioned by some authors [105],
I quoted here only the results of  direct specific-heat
measurements.

\bigskip
\noindent
{\it Exponent $\beta $\/}
\par\medskip
The exponent $\beta $ has been measured by  neutron,
X-ray  and M\"osbauer techniques.
While all authors agreed in that the transition was continuous, 
the reported
values of  $\beta $ were scattered wildly as
0.21 (Tb; X-ray), 0.23
(Tb; neutron), 0.25(Tb; neutron), 0.3(Ho; neutron), 
0.335(Dy; M\"osbauer), 
0.37(Ho; X-ray), 0.38(Dy; neutron), 
0.39(Ho; neutron) to 0.39(Dy; neutron).
It is not easy to read off a systematic tendency from this.
Some of the values, particularly $\beta $
for Tb, were
close to the $n=2$
chiral value, but other values, especially those 
obtained by  neutron and X-ray diffraction for Ho and Dy 
tend to give much larger 
values  close to the $O(4)$ value.
\par\medskip
\noindent
{\it Exponent $\gamma $ and $\nu $\/}
\par\medskip
The exponents $\gamma $ and $\nu $ 
have been measured by neutron and X-ray
scatterings.  Neutron-scattering
measurements by Gaulin, Hagen and Child gave
$\gamma =1.14\pm 0.04$ 
$\nu =0.57\pm 0.04$ for Ho, and
$\gamma =1.05\pm 0.07$,  $\nu =0.57\pm 0.05$ for Dy, 
which were close to the 
$n=2$ chiral values [108]. 
More recent X-ray and high-precision neutron-scattering studies
on Ho by Thurston and coworkers 
revealed interesting new features [109]. Critical scattering 
above $T_N$ actually consisted of two components characterized by
mutually different exponents: 
A broad component characterized by the exponents
$\nu =0.55\pm 0.04$ and $\gamma =1.24\pm 0.15$,
which was associated with 
the bulk contribution inside the sample, and   
a narrow component characterized by 
the exponents $\nu =1.0\pm 0.3$ and $\gamma =3.4\sim 4.5$,  
which  came from
the skin part of
the sample.
High-precision neutron-scattering for Tb
also established the existence of such two length scales [110].
Exponents associated with the
broad component were in agreement with the earlier measurements.
Exponents associated with the narrow component
was explained by Altarelli {\it et al\/} [111]
as governed by the long-range disorder fixed point [112], on the
assumption that the skin layer of Ho
contains a number of edge-dislocation
dipoles.
Anyway, these experiments have clearly shown that,
in order to get the bulk critical properties from
the measurements  sensitive to the defected skin layer, 
special care has to be taken to extract 
the bulk component from the signal.
\par\bigskip
\noindent
{\it First-order transition?\/}
\par\medskip
As mentioned, a few authors claimed that their experimental data
for Ho and Dy were suggestive of
a weak first-order transition [11,12].
Probably,  first experimental 
claim that the transition in Ho might be weakly
first order was made by Tindal, Steinitz and Plumer
based on their thermal expansion 
measurements of Ho along the $a$ axis [11]. 
These authors observed a jump-like anomaly
in the thermal expansivity along the $a$ axis, although
no such anomaly was detected along the $c$ axis. Tindall {\it et al\/}
interpreted this anomaly
as an evidence of a first-order transition. 
Later thermal-expansion  measurements by White on Ho
along the $a$-axis,  however, 
lead to the opposite conclusion that the transition was continuous
[113],
and the situation  remains unclear.
Putting aside
such discrepancy among independent measurements,
an apparent jump-like behavior observed by Tindal {\it et al\/}
appears to be explained by
the standard power-law singularity characteristic
of a continuous transition of the
form,
$$\Delta a/a\approx b_0+b_1t+c_\pm \mid t \mid ^{1- \tilde \alpha },
\ \ \ t\equiv (T-T_N)/T_N,$$
{\it if\/} $b_0>0$, $b_1>0$, 
$c_+<0$ {\it and\/} $c_->0$,
as long as the 
exponent $\tilde \alpha $, usually identified as the specific-heat
exponent $\alpha $, is positive.
Note that the coefficients $c_\pm $
could be negative even if the total
thermal expansivity is to be positive. 
Hence,  the data of Ref.11
cannot be regarded as an unequivocal proof of  first-order
transition.

While earlier thermal-expansion  measurements on  
rare-earth metals Dy and Tb 
observed a continuous transition [114],  
Zachowski {\it et al\/} suggested
that the paramagnetic-helimagnetic transition 
of Dy might also be first order
based on their
observation of  deviation from a
single power-law scaling behavior  in an
immediate vicinity of $T_N$ [12]. Care has to be taken in this
interpretation, however, since
apparent deviation from the scaling behavior 
in a  vicinity of $T_N$ could arise
from many secondary effects, such as 
rounding  due to
impurities or inhomogeneities, insufficiency of 
temperature control, 
crossover of yet unidentified 
nature, or even the contribution from the defected skin layer,  
{\it etc.\/} Therefore,  
in order to experimentally conclude 
that the transition is really first order,
one should give a reliable lower bound to the discontinuity
of some physical quantity at the transition, 
such as finite latent heat.
At  present,
there appears to be 
no such firm experimental evidence of  first-order 
transition.
\par\medskip\noindent
{\it Discussion\/}
\par\medskip
As shown, experimental data of rare-earth metals  
are sometimes mutually conflicting. 
Below,
I wish to try to discuss 
the possible cause  of the
conflict together with its possible resolution.

One  point to be remembered is
that the magnetic interaction in these rare-earth
metals is the long-range RKKY interaction whose range is of order
the pitch of the helix. It 
means that, when one is  far away from
$T_N$ and the correlation length is smaller than the helix pitch, 
one should have the ordinary mean-field critical
behavior characterized by $\alpha =0$, $\beta =0.5$ and $\gamma =1$
{\it etc\/} [26]. 
Only when one further approaches $T_N$ and the correlation 
length gets
longer, one should have a true
asymptotic critical behavior.  
If one assumes
that the asymptotic critical
behavior 
is also of $n=2$ chiral universality,
one expects a mean-field to $n=2$ chiral crossover,
$\alpha =0\rightarrow 0.34$, $\beta =0.5\rightarrow 0.25$,  
$\gamma =1\rightarrow 1.13$ {\it etc\/}.
In fact, this scenario appears to account for many of the 
experimental results. For example,
earlier specific-heat measurements by Ledermann and Salamon [7]
where the data  exhibited
a crossover from a smaller $\alpha $ value  to a larger $\alpha $ 
value appears consistent with this scenario.
In case of $\gamma $, 
since the mean-field value $\gamma =1$ and the $n=2$ chiral value
$\gamma \simeq 1.13$ happen to be rather close, 
this crossover would  be hard to  detect clearly, which is also
consistent with  experiment [108].
\par
Another important ingredient might be the possible contribution from
the defected skin part of  sample as discussed above.
While the contribution from the skin part is separable above $T_N$
by analyzing the lineshape of the scattering function [109,110], 
such separation is not straightforward 
below $T_N$  since both the bulk
and the skin contributions yield resolution-limited Bragg peaks.
This means that 
the Bragg intensity observed so far is likely to be a
superposition of these two distinct  components, 
each with different
exponents $\beta $.
If one assumes  the above
scenario, the bulk component exhibits a
crossover from $\beta =0.5$ (ordinary mean-field) to $\beta \simeq
0.25$ ($n=2$ chiral), while, according to Ref.111, 
the skin component exhibits a
behavior governed by the  long-range disorder fixed point 
characterized by $\beta =0.5$.
So, rather complicated situation might indeed occur in rare-earth
metals, and
special care has to be taken in extracting  information
about the asymptotic bulk critical
behavior.
To my knowledge, experimental analysis fully taking account of such
complication
has not yet been done
especially below $T_N$.
Thus, it is highly desirable to
extract the bulk component {\it below\/} $T_N$ by separating the 
contribution of the skin
component by some experimental device.
\par
One possible experiment to bypass the above complications 
might be to study 
{\it insulating\/} helimagnets.
There is at least one candidate material, VF$_2$, which is known
to exhibit a paramagnetic-helimagnetic transition [115].
Since the magnetic interaction in VF$_2$ is {\it short-ranged\/}, 
one need not worry about the slow crossover from the mean-field
behavior, and 
hopefully, the effect of defected skin part would be less severe.
If so,  information about the critical properties
of VF$_2$ would be  valuable to disentangle the
present complicated situation concerning helimagnets, 
and I wish to urge experimentalists
to try such experiments.
\par
So, one  plausible scenario proposed here  is 
that the asymptotic criticality
of helimagnets is also 
of $n=2$ chiral universality as in the case of
stacked-triangular antiferromagnets, which is 
blurred and masked
by the slow crossover from the ordinary
mean-field behavior caused by
the long-range  RKKY interaction as well as by the
contribution of the defected skin part of  sample.
Of course, this hypothesis should be tested by  experiments, 
some
of which have been proposed above.
\par
\bigskip
\noindent
(c) {\it Measurements of chirality\/}
\par\medskip
Chirality is a quantity playing an 
important role  
in the noncollinear transitions. Hence,  it is of great
interest to experimentally measure the chirality. Since the chirality
is a multispin variable of higher order in the original spin
variables, its direct experimental detection needs some
ingenuity. Plumer, Kawamura and Caill\'e pointed out that, 
if one could
prepare a sample with a single chiral domain, the average
total chirality  $\bar \kappa $ could be measured
by using polarized neutrons [116]. These authors also suggested that
a single chiral domain might be prepared by cooling the sample
under applied electric fields. Experimental attempt 
along this direction was made by Visser {\it et al\/} [117]. 
Maleyev suggested that the chirality might be observable by measuring
the polarization dependent part of  neutron scattering in applied
magnetic fields [118]. 
Federov {\it et al\/} suggested that the chirality
sense might be controlled by applying the elastic torsion, which could
be used to prepare a single chiral-domain sample [119]. 
To the author's knowledge, however, these methods and ideas have
not yet be fully 
substantiated. Direct experimental detection of chirality
is certainly a challenging problem,
which may serve to provide  new experimental
tool to look into the noncollinear orderings.
\par
\bigskip
\bigskip
\noindent
{\bf \S 7. Critical and 
multicritical behaviors under magnetic fields}
\par
\medskip
In this section, I will 
review the phase transition of stacked-triangular
antiferromagnets under applied magnetic fields.
Let us first begin with the case of
unfrustrated  collinear 
antiferromagnets on bipartite lattices. 
Typical magnetic field - temperature phase diagrams of such
weakly anisotropic antiferromagnets
are illustrated in Fig.15 for the cases of axial (Ising-like) 
anisotropy
with  field applied along  an easy axis (a), 
and for the case
of planar ({\it XY\/}-like) anisotropy with  field applied in an
easy plane (b). 
Axial magnets in a field
exhibit a multicritical point, termed bicritical point, at which
two  critical lines and a first-order spin-flop line meet: See
Fig.15a.
Critical properties of these axial magnets 
along the critical lines and
at the bicritical point were theoretically studied by 
Fisher and Nelson [120], and by Kosterlitz,
Nelson and Fisher [121], with the results given in Fig.15.  The
criticalities
are of standard $O(n)$ universality
with $n=1,2,3$. Applying a scaling theory, Fisher {\it et al\/}
derived various predictions, which were supported by subsequent 
experiments [122]. It thus appears that the critical and
the multicritical behaviors of
unfrustrated collinear antiferromagnets in a field are now fairly
well understood.

In the case of  frustrated
noncollinear antiferromagnets such as stacked-triangular
antiferromagnets, typical magnetic phase diagrams
are shown in Fig.16 for the cases of axial (Ising-like) anisotropy
with  field applied along an easy axis (a),
and for the case of planar ({\it XY\/}-like) anisotropy 
with  field applied in an
easy plane (b).
In the axial case, three critical lines and a first-order spin-flop
line meet at a new type of multicritical point at ($T_m$, $H_m$):
See Fig.16a. In the planar case, two distinct critical lines meet
at a zero-field multicritical point, termed tetracritical point:
See Fig.16b. 

Such novel features of the phase diagrams 
and the multicritical behaviors of stacked-triangular antiferromagnets
were first observed experimentally.
In the axial case, phase diagram 
with a novel multicritical point was found by
Johnson, Rayne and Friedberg in 1979 for 
CsNiCl$_3$ by means of  susceptibility measurements [38], while
in the planar case phase diagram
with a zero-field tetracritical point was
determined by Gaulin {\it et al\/} 
in 1989 for
CsMnBr$_3$ by means of neutron-scattering measurements [39].
Subsequent phenomenological free-energy analysis successfully
reproduced the main qualitatative features of these phase diagrams
[123,124].
These multicritical behaviors
in a field were also reproduced
by subsequent Monte Carlo simulations [125-127].

Scaling  analysis of the critical and the multicritical
properties of stacked-triangular antiferromagnets 
under  magnetic fields
was made by Kawamura, Caill\'e and Plumer based on the 
chiral-universality scenario [40,79]: According to this scaling theory,
in the axial case, the criticality along the two low-field 
critical lines is of  standard $XY$ universality, while
the one  along the high-field  critical line is  of  
$n=2$ chiral universality. Meanwhile, 
the multicritical behavior right at the
multicritical point is predicted to be of $n=3$ chiral universality.
In the planar case, the criticality along the
higher-temperature critical line 
is  of  {\it XY\/} universality, while the one along
the lower-temperature critical line
is of   Ising universality. 
The multicritical (tetracritical) behavior  at the zero-field
transition point is of $n=2$ chiral universality governed by the
$n=2$ chiral fixed point.

Scaling theory further predicted that,
in the axial case, three critical lines should merge at
the multicritical point tangentially with the first-order spin-flop
line as [40]
$$\mid H-H_m\mid \propto \mid T-T_m\mid ^\phi,\eqno(7.1)$$
where the exponent 
$\phi \sim 1.06$ is common among the three critical lines. 
In fact, $\phi $ is  the
anisotropy-crossover exponent at the $n=3$ chiral fixed point
identified in the RG analysis in \S 4(b). 

Similarly, in the planar case, it is
predicted that the two critical lines in external fields
should merge at the
zero-field tetracritical point as [40]
$$H^2 \propto \mid T-T_m\mid ^\phi,\eqno (7.2)$$ 
where $\phi \sim 1.04$ is the 
anisotropy-crossover exponent at the $n=2$ chiral fixed point,
common between the two critical lines.
Near the tetracritical point,
the zero-field uniform susceptibility  was 
predicted to behave as [40,42,79]
$$\chi (T,H=0)\approx C_{\pm }\mid T-T_m\mid ^
{-\tilde \gamma}+[{\rm 
less\ singular\ and\ regular\ parts}],\eqno (7.3)$$ 
where $\tilde \gamma =-(2-\alpha -\phi)\sim -0.56$.

These scaling predictions were tested by subsequent experiments.
In the axial case,  criticality along the three critical
lines as well as at the multicritical point were examined by
several authors.
In particular, the predicted
$n=2$ chiral behavior along the high-field critical line as well as
the $n=3$ chiral behavior at the multicritical point were very well
confirmed by  specific-heat measurements by Beckmann, Wosnitza and 
von L\"ohneysen on CsNiCl$_3$ [91],
by  birefringence measurements by Enderle, Furtuna and 
Steiner on CsNiCl$_3$ and CsMnI$_3$ [92], 
and by  neutron-diffraction measurements  by 
Enderle, Schneider, Matsuoka and Kakurai on CsNiCl$_3$ [93].
The behavior of the phase boundaries 
near the multicritical point
was  investigated by Poirier {\it et al\/} for CsNiCl$_3$, who found
by means of  ultrasonic velocity measurements 
that the low-temperature low-field critical line between the
collinear and noncollinear phases (regions 2 and 3 in Fig.17) 
exhibited a
`turnover' in a close vicinity of the multicritical point
to merge into the first-order spin-flop line, as shown in Fig.17
[128].
This turnover behavior was not expected by the mean-field theory,
but in accord with the scaling prediction.
Katori, Goto and Ajiro [129], and Asano {\it et al\/} [130] determined 
by magnetization
measurements the phase diagrams of other
axial stacked-triangular antiferromagnets  CsNiBr$_3$ and  
CsMnI$_3$, and emphasized  universal aspects of the
phase diagrams.

Along the two low-field critical lines, theory expects the standard {\it XY\/}
critical behavior. Experimentally, 
the critical properties at these two transition points were 
studied in zero
field  by several methods, 
including NMR [131],  
neutron scattering
[132]  for CsNiCl$_3$,  neutron scattering [133,134] 
and specific heat [135] for CsMnI$_3$. Most of the results are
consistent with the expected {\it XY\/} criticality, 
although significant
deviation was observed in a few cases such as the exponents
$\gamma $ and $\nu $ reported in Ref.134. A part of such deviation
may be ascribed to the proximity effect of the $n=3$ chiral
behavior realized at the multicritical point at $H=H_m$.

In the planar case, the situation is not
entirely satisfactory. 
Concerning the behavior of the two
critical lines near the zero-field tetracritical point, 
Gaulin {\it et al\/} reported by neutron scattering for CsMnBr$_3$  
the crossover exponents
$\phi _{{\rm P-II}}\sim 1.21$ and $\phi 
_{{\rm II-I}}\sim 0.75$ for the high-
and low-temperature critical lines, respectively, which differed
considerably  from the scaling results, 
$\phi _{{\rm P-II}}= \phi _{{\rm II-I}}\sim 1$.
Reanalysis of the data
by Gaulin, however,  revealed that, once the uncertainty of $T_m$ 
was taken into account in the analysis, 
the experimental data were not 
inconsistent with the scaling results [42]. Goto, Inami and
Ajiro  found by magnetization measurements
$\phi _{{\rm P-II}}= 1.02 \pm 0.05$ and $\phi _{{\rm II-I}}= 
1.07\pm 0.05$ 
for CsMnBr$_3$ [136], which were
in good agreement with the theoretical values.
By contrast, markedly smaller values, 
$\phi _{{\rm P-II}}= 0.78\pm 0.06$ and $\phi _{{\rm II-I}}= 0.79
\pm 0.06$, 
were reported  by Tanaka, Nakano and Matsuo for 
the {\it XY\/} stacked-triangular antiferromagnet CsVBr$_3$ by
susceptibility measurements [137], while the values, 
$\phi _{{\rm P-II}}=0.76\pm 0.1 $ and $\phi _{{\rm II-I}}
=0.81\pm 0.1$, were reported 
by Weber, Beckmann, Wosnitza and
von L\"ohneysen for CsMnBr$_3$ by 
specific-heat measurements [138]. 
The cause of this discrepancy
is not clear. From theoretical side, although the prediction that
the  exponent $\phi $ is common among the critical lines is a
direct consequence of the chiral-universality picture, its precise
value is still subject to  large uncertainties, because
it has not yet been determined by  reliable numerical methods
such as  extensive  Monte Carlo simulation.
It is thus desirable
to give a more reliable numerical estimate of the anisotropy-crossover
exponent $\phi $.

It turns out that the zero-field transition point of 
RbMnBr$_3$ is also a
tetracritical point in the magnetic field -- 
temperature phase diagram [139-141].
The associated crossover exponents were determined by 
Heller {\it et al\/} by means of neutron scattering
as $1.00\pm 0.35$ and $1.07\pm 0.25$, for
the higher-temperature and the lower-temperature critical lines,
respectively [141].

The zero-field susceptibility of CsMnBr$_3$ was measured by 
Mason, Stager, Gaulin and Collins [142].
These authors interpreted
their data as being inconsistent with
the scaling prediction 
on the assumption that
the coefficients
of the leading singularity, $C_{\pm }$ in eq.(7.3), 
were both positive and that the  contribution from the
regular and  less singular terms were zero.
However, once one properly takes account of
the fact that the sign of $C_{\pm }$ could be different on both sides
of $T_m$ and that there generally exists a finite 
contribution 
from the regular and less singular terms, the experimental data are
consistent with the scaling theory [42,79]. 
\bigskip
\bigskip
\noindent
{\bf \S 7. Summary}
\par
\medskip
Recent theoretical and experimental studies on  phase transitions
of  noncollinear or canted magnets, including both
stacked-triangular antiferromagnets and helimagnets, were reviewed
with particular emphasis on the novel critical and multicritical 
behaviors observed
in these magnets.
\par
Theoretical analyses based on various
renormalization-group techniques, which usually gave good 
results for standard unfrustrated magnets, have given
somewhat inconclusive
and sometimes conflicting 
results
concerning the nature of  the noncollinear transitions.
Special care appears to be necessary
in applying the standard  RG methods
to the system
with  nontrivial structure in the order-parameter
space as in the present problem.  Nevertheless,
as was discussed in detail in \S 4, 
most plausible possibility suggested from the RG analyses is 
either the transition is
continuous governed by a new fixed point 
(chiral universality),
or else, the transition is first order.

Most of the recent extensive Monte Carlo simulations performed on 
{\it XY\/} and Heisenberg stacked-triangular antiferromagnets
suggest the occurrence of a continuous transition
characterized by the  exponents  significantly different
from the standard $O(n)$ exponents. In that sense, these Monte Carlo
results support the
chiral-universality scenario. 
the bulk of various experiments on 
stacked-triangular {\it XY\/} and Heisenberg antiferromagnets
have also yielded results in favor of the chiral-universality
scenario:
A continuous transition characterized by the
novel exponents close to those
obtained by Monte Carlo simulations has been observed.
Meanwhile, in the case of {\it XY\/} spins, 
many of the Monte Carlo and 
experimental results also appear to be marginally consistent with the
mean-field tricritical behavior, while such behavior
is not suggested by some of the data such as the specific-heat
exponent, specific-heat amplitude ratio  and  chirality exponents. 
From a theoretical viewpoint, the mean-field tricritical
behavior dictated by the trivial Gaussian fixed point is  rather
unlikely even when the system happens to be
just at its tricriticality, as long 
as the generic noncollinear criticality is  {\it not\/} of the
standard $O(n)$ universality.
Thus, 
at least in the case of 
stacked-triangular antiferromagnets,
there appears to be reasonable  evidence both from Monte Carlo 
simulations and experiments that a
new chiral universality class is in fact realized. 

There still seems to exist a slight 
chance of a  weak first-order transition, though, either
from Monte Carlo
simulations or from experiments. 
Although this point is to be examined, it seems already clear from
recent extensive studies  that there exists a rather wide and
well-defined critical regime, say at $10^{-1}>t>10^{-3}$,
characterized by a set of novel critical exponents and
amplitude ratio, {\it which are 
universal among various noncollinear magnetic 
materials and model systems\/}.
This observation strongly suggests {\it the existence of an underlying
novel fixed point governing the
noncollinear criticality\/}. The remaining possibility is that this
fixed point may be slightly complex-valued.
It is certainly interesting to further examine
the order of the transition
both from careful numerical simulations and 
from high-precision experiments,
either to get an unambiguous evidence of 
first-order transition or to push the 
limit of  continuous nature of the transition further away.
To do this experimentally, one needs to choose appropriate
materials which do not have a weak 
perturbative interaction which breaks
the chiral symmetry. In addition,
to be sure that the transition is first order, one should give a
reliable lower bound on the discontinuity of some physical
quantities such as the latent heat. Mere observation of 
deviation from a simple power-law behavior in a close vicinity 
of $T_N$
is not quite enough to conclude that the transition is first order,
since such deviation could arise from many secondary effects.
Also one should recognize that, even when there exists a well-defined
chiral universality class, 
it is completely  possible that some  systems sharing the
same chiral symmetry exhibit first-order transition due to
the difference in
nonuniversal details of certain microscopic parameters.

In contrast to stacked-triangular antiferromagnets, 
the present situation in helimagnets (spiral magnets) is
less clear. In particular,  experimental situation for rare-earth
helimagnets has been  confused for  years now. I have
proposed one possible
scenario to solve this confusion based on the chiral-universality
scenario, where the combined effects of the 
long-range nature of the RKKY interaction and the contribution from
the defected skin part
hinder the observation of an ideal chiral critical behavior. 
It might be interesting to test the proposal by further experiments.
On numerical side, it might be interesting to perform further
Monte Carlo simulations on helimagnets by 
paying attention to the effects of
boundary conditions.

To sum up, phase transitions of frustrated noncollinear magnets
exhibit novel behaviors  different from  standard 
unfrustrated collinear magnets. Although there is not a complete
consensus among researchers, many of both 
experimental and numerical results on stacked-triangular
antiferromagnets point to
the occurrence of phase transitions of  new chiral universality class,
distinct from the standard $O(n)$ Wilson-Fisher universality class.
As a reflection of  richer 
structure of its order parameter, the noncollinear transitions
also possess some unique physical quantities such as  chirality  
which have no counterpart in the standard unfrustrated magnets.
Such rich inner symmetry also leads to  unique
magnetic phase diagrams in external fields 
with  novel multicritical behaviors.
In this decade, there has been a stimulating and fruitful interplay
between theory and experiment in the area.
Hopefully, further theoretical as well as experimental works
will clarify novel features of the noncollinear
transitions, which might
serve to enlarge and deepen our  understanding of 
phase transitions and critical phenomena.
\par

The author is thankful to Prof. D.P. Belanger for a reading of the
manuscript.
\bigskip
\vfil\eject

\noindent
{\bf References}
\medskip\noindent
\item{[1]} Villain J 1977 Phys. C{\bf 10} 4793 
\item{[2]} Miyashita S and  Shiba H 1985  J. Phys. Soc. Jpn. 
{\bf 53} 1145
\item{[3]}  Kawamura H and  Miyashita S 1984 J. Phys. Soc. Jpn. 
{\bf 53}  4138
\item{[4]}  Bak P and  Mukamel D 1976 Phys. Rev. B{\bf 13}
5086
\item{[5]}  Garel T and  Pheuty P 1976 J. Phys. C{\bf 9}
L245
\item{[6]}  Eckert J and  Shirane G 1976 Solid State Comm. {\bf 19} 
911
\item{[7]}  Lederman E L and  Salamon M B  1974
Solid State Comm. {\bf 15}  1373
\item{[8]}  Loh E, Chien C L and Walker J C 1974
Phys. Letters A{\bf 49}  357
\item{[9]} Dietrich O W  and  Als-Nielsen J 1967
Phys. Rev. {\bf 162} 315
\item{[10]}  Barak Z and Walker M B 1982  Phys. Rev. B{\bf 25}
1969
\item{[11]}  Tindall D A, Steinitz M O   and  Plumer M L 1977 
J. Phys. F{\bf 7} L263
\item{[12]} Zachowski S W, Tindall D A,  Kahrizi M,  Genossar J
and  Steinitz M O 1986 J. Magn. Magn. Mater. {\bf 54}-{\bf 57}
707
\item{[13]}  Tang C C, Stirling W G, Jones D L,  Wilson C C,
Haycock P W,  Rollason A J,  Thomas A H and  Fort D 1992
J. Magn. Magn. Mater. {\bf 103}  86
\item{[14]}  Tang C C, Haycock P W, Stirling W G, Wilson C C,
Keen D and  Fort D 1995
Physica B{\bf 205}  105
\item{[15]}  Thurston T R,  Helgesen G, Hill J P,  Gibbs D,
Gaulin B D and  Simpson P J 1994
Phys. Rev. B{\bf 49}  15730
\item{[16]}   Du Plesis P, Venter A M and Brits G H F 1995
J. Phys.  Condens. Matter {\bf 7}  9863
\item{[17]}   Du Plesis P, van Doorn C F  and van Delden D C 1983
J. Magn. Magn. Mater.  {\bf 40}  91
\item{[18]}  Brits G H F and  Du Plesis P 1988 
J. Phys.  F{\bf 18}  2659
\item{[19]}  Kawamura H 1985 J. Phys. Soc. Jpn. {\bf 54}  3220; 
1987 J. Phys. Soc. Jpn. {\bf 56} 474
\item{[20]}  Kawamura H  1986 J. Phys. Soc. Jpn. {\bf 55} 2095; 
1989 J. Phys. Soc. Jpn. {\bf 58} 584
\item{[21]}  Kawamura H 1992 J. Phys. Soc. Jpn. {\bf 61}  1299
\item{[22]}  Kawamura H 1988 J. Appl. Phys. {\bf 63}  3086
\item{[23]}  Kawamura H 1988 Phys. Rev. B{\bf 38}  4916; 1990
Phys. Rev. B{\bf 42}
2610(E)
\item{[24]}  Mason T E, Collins M F  and  Gaulin B D 1987
J. Phys. C{\bf 20}  L945; 1989
Phys. Rev. B{\bf 39}  586
\item{[25]}  Ajiro Y,  Nakashima T,  Unno Y,  Kadowaki H,  Mekata M
and  Achiwa N 1988 J. Phys. Soc. Jpn. {\bf 57}  2648;
Kadowaki H,  Shapiro S M,  Inami T and Ajiro Y 1988
J. Phys. Soc. Jpn. {\bf 57}  2640
\item{[26]}  Wang J,  Belanger D P and  Gaulin B D 1991
Phys. Rev. Lett. {\bf 66}  3195
\item{[27]}  Deutschmann R, L\"ohneysen H v,  Wosnitza J, 
Kremer R K and  Visser D 1992 Europhys. Lett. {\bf 17}  637
\item{[28]}  Takeda K,  Ury\^u N,  Ubukoshi K and  Hirakawa K 1986
J. Phys. Soc. Jpn. {\bf 55}  727
\item{[29]}  Kadowaki H,  Ubukoshi K,  Hirakawa K, Martin\'ez J L  
and  Shirane G 1988 J. Phys. Soc. Jpn. {\bf 56}  4027
\item{[30]}  Wosnitza J,   Deutschmann R, L\"ohneysen H v and 
Kremer R K 1994
J. Phys. Condens. Matter {\bf 6}  8045
\item{[31]}  Azaria P,  Delamotte B and Jolicoeur Th 1990
Phys. Rev. Lett. {\bf 64}  3175
\item{[32]}  Azaria  P,  Delamotte B,  Delduc F and Jolicoeur Th 1993
Nucl. Phys. B{\bf 408}  485
\item{[33]}  Bhattacharya T,  Billoire A,  Lacaze R and Jolicoeur Th
1994 J. Phys. I {\it France} {\bf 4}  181
\item{[34]} Boubcheur E H,  Loison D and Diep H T 1996
Phys. Rev. B{\bf 54}  4165
\item{[35]}   Loison D and  Diep H T 1994
Phys. Rev. B{\bf 50}  16453
\item{[36]}  Plumer M L and  Mailhot A 1994 
Phys. Rev. B{\bf 50}  16113
\item{[37]}  Mailhot A, Plumer M L and  Caill\'e A 1994
Phys. Rev. B{\bf 50}  6854
\item{[38]}   Johnson P B,  Rayne J A and  Friedberg S A 1979 
J. Appl. Phys. {\bf 50}  583
\item{[39]}   Gaulin B D,  Mason T E,  Collins M F and  Larese J Z 
1989 Phys. Rev. Lett. {\bf 62}  1380
\item{[40]}  Kawamura H,  Caill\'e A and  Plumer M L 1990
Phys. Rev. B{\bf 41}  4416
\item{[41]}  Kawamura H 1992 in {\it Recent Advances in Magnetism 
of Transition Metal 
Compounds\/}, edited by A. Kotani and N. Suzuki (World Scientific,
Singapore) p.335
\item{[42]}  Gaulin B D 1994 in 
{\it Magnetic Systems with Competing Interactions
}  
edited by H.T. Diep (World Scientific,
Singapore) p.286
\item{[43]}  Plumer M L,  Caill\'e A,  Mailhot A and  Diep H T, 1994
in {\it Magnetic Systems with Competing Interactions} 
edited by H.T. Diep (World Scientific,
Singapore) p.1
\item{[44]} Collins M F and  Petrenko O A 1997 
Can. J. Phys. {\bf 75}  605
\item{[45]}   Kawamura H 1990
Prog. Theor. Phys. Suppl. {\bf 101}  545
\item{[46]}  Zhang W, Saslow W M  and  Gabay M 1991 
Phys. Rev. B{\bf 44}  5129;
Zhang W,  Saslow W M,  Gabay M and  Benakli M 1993
Phys. Rev. B{\bf 48}  10204
\item{[47]}  Jones D R T,  Love A and  Moore M A 1976  
J. Phys. C{\bf 9} 743
\item{[48]}  Bailin D,  Love A and Moore M A 1977  J. Phys. C{\bf 10}
1159
\item{[49]}  Joynt R 1991  Europhys. Lett.
{\bf 16}  289; 1993 Phys. Rev. Lett. {\bf 71}  3015
\item{[50]}  Granato E and  Kosterlitz J M 1990  Phys. Rev. Lett.
{\bf 65}  1267
\item{[51]}  Toulouse G and  Kl\'eman M 1976 J. de Phys. Lett.  
{\bf 37}  149; Mermin N D 1979  Rev. Mod. Phys. {\bf 51}
591
\item{[52]}  Kosterlitz J M and  Thouless D J 1973 J. Phys. C{\bf 6}
1181;  Kosterlitz J M 1974 C{\bf 7}  1046
\item{[53]}  Lee D H,  Jonnopoulos J D,  Negele J W and
Landau D P 1984 Phy. Rev. Lett. {\bf 52}  433;  
1986 Phys. Rev. B{\bf 33}  450
\item{[54]}  Teitel S and  Jayaprakash C 1983 
Phys. Rev. B{\bf 27} 
598
\item{[55]}  Ramirez-Santiago G and  Jos\'e J V 1992
Phys. Rev. Lett. {\bf 68}  1224; 1994 Phys. Rev. {\bf 49}  
9567 and references therein
\item{[56]}  Olsson P 1995
Phys.  Rev. Lett. {\bf 75}  2758 and references therein
\item{[57]}  Wintel M,  Everts H U and  Apel W 1995
Phys. Rev. B{\bf 52}  13480
\item{[58]}  Kawamura H and  Kikuchi M 1993 Phys. Rev. B{\bf 47} 
1134
\item{[59]}  Southern B W and  Xu H-J 1995
Phys. Rev. B{\bf 52}  R3836
\item{[60]}  Kohring G,  Shrock R E and  Wills P 1986 
Phys. Rev. Lett.
{\bf 57}  1358; 
Williams G 1987 Phys. Rev. Lett.
{\bf 59}  1926;  Shenoy S R 1989 
Phys. Rev. B{\bf 40}  7212
\item{[61]}  Antonenko S A,  Sokolov A I and  Varnashev K B 1995
Phys. Lett. A{\bf 208}  161
\item{[62]}  Mudrov A I and  Varnashev K B 
1997 cond-mat/9712007; 1998 cond-mat/9802064.
\item{[63]}  Antonenko S A and  Sokolov A I 1994
Phys. Rev. B{\bf 49}  15901
\item{[64]}  Cowley R A and  Bruce A D 1978
J. Phys. C{\bf 11}  3577
\item{[65]}  Shpot N A 1989
Phys. Lett. {\bf 142}  474
\item{[66]}  Kawamura H unpublished
\item{[67]}  Kawamura H 1988
Phys. Rev. B{\bf 38}  960
\item{[68]}  Jolicoeur Th 1995 Europhys. Lett. {\bf 30}  555
\item{[69]}  Halperin B I,  Lubensky T C and  Ma S K 1974
Phys. Rev. Lett. {\bf 32}  292
\item{[70]}  Dasgupta C and  Halperin B I 1981
Phys. Rev. Lett. {\bf 47}  1556
\item{[71]}  Olsson P and  Teitel S
1997 cond-mat/9710200 and references therein
\item{[72]}  Radzihovsky L 1995
Europhys. Lett. {\bf 29}  227
\item{[73]}  Kleinert H,  Thomas S and Schote-Frohline V 1996 
quant-ph/9611050 and references therein
\item{[74]}  Kawamura H  unpublished
\item{[75]}  Kunz H and  Zumbach G 1993
J. Phys. A{\bf 26}  3121
\item{[76]}  Dobry A and  Diep H T 1995 
Phys. Rev. {\bf 51}  6731
\item{[77]}   Zumbach G 1993
Phys. Rev. Lett. {\bf 71}  2421
\item{[78]}   Zumbach G 1994
Phys. Lett. A{\bf 190}  225; 1994 Nucl. Phys. B{\bf 413}  771
\item{[79]}  Kawamura H 1993 
Phys. Rev. B{\bf 47}  3415
\item{[80]}   Kawamura H 1991
J. Phys. Soc. Jpn. {\bf 60}  1839
\item{[81]}   Zumbach G 1995
Nucl. Phys. B{\bf 435}  753
\item{[82]}   David F and  Jolicoeur Th 1996
Phys. Rev. Lett. {\bf 76}  3148
\item{[83]}  See, for example,  Castilla G E and  Chakravarty S 1993
Phys. Rev. Lett. {\bf 71}  384; 1997 Nucl. Phys. B{\bf 485}
613 and references therein
\item{[84]}   Kawamura H 1990
J. Phys. Soc. Jpn. {\bf 59}  2305
\item{[85]}   Saul L 1992
Phys. Rev. B{\bf 46}  13847
\item{[86]}   Reimers J N,  Greedan J E and  Bj\"orgvinsson M 1992
Phys. Rev. B{\bf 45}  7295
\item{[87]}   Mailhot A and  Plumer M L 1993
Phys. Rev. B{\bf 48}  9881
\item{[88]}   Plumer M L and  Mailhot A 1997
J. Phys. Condens. Matter  {\bf 9}  L165
\item{[89]}   Diep H T 1989
Phys. Rev.  B{\bf 39}  397
\item{[90]}   Saslow W M,  Gabay M and  Zhang W  1992
Phys. Rev. Lett. {\bf 68}  3627
\item{[91]}   Beckmann D,  Wosnitza J and  L\"ohneysen H v 1993
Phys. Rev. Lett.  {\bf 71}  2829
\item{[92]}   Enderle M,  Furtuna G and  Steiner M 1994
J. Phys. Condens. Matter  {\bf 6}  L385
\item{[93]}   Enderle M,  Schneider R,  Matsuoka Y and  Kakurai K
1997
Physica  B{\bf 234}-{\bf 236}  554
\item{[94]}   Yelon W B and  Cox D E 1972 Phys. Rev. B{\bf 6}
204
\item{[95]}   Oohara Y,  Kadowaki H and  Iio K 1991
J. Phys. Soc. Jpn. {\bf 60}  393
\item{[96]}   Oohara Y,  Iio K and  Tanaka H 1991
J. Phys. Soc. Jpn.  {\bf 60}  4280
\item{[97]}   Eibsh\"utz E,  Sherwood R,  Hsu F L and  Cox D E 1972
AIP Conf. Proc. {\bf 17}  684
\item{[98]}   Kato T,  Iio K,  Hoshino T,  Mitsui T and  Tanaka H
1992 J. Phys. Soc. Jpn.  {\bf 61}  275
\item{[99]}   Kato T,  Asano T,  Ajiro Y,  Kawano S,  Ishii T
and  Iio K 1995
Physica,  B{\bf 213} \& {\bf 214}  182
\item{[100]}   Tanaka H,  Iio K and  Nagata K 1985
J. Phys. Soc. Jpn.  {\bf 54}  4345
\item{[101]}   Mekata M,  Ajiro Y,  Sugino T,  Oohara A,  Ohara K,
Yasuda S,  Oohara S and  Yoshizawa H 1995
J. Magn. Magn. Mater.  {\bf 140} - {\bf 144}  38
\item{[102]}   Schotte U,  Stuesser N,  Schotte K D,
Weinfurter H,  Mayer H M and 
Winkelmann M 1994 J. Phys. Condens. Matter {\bf 6}  10105;
Stuesser N, Schotte U,  Schotte K D and  Hu X 1995 Physica
B{\bf 213} \& {\bf 214}  164
\item{[103]}   Weber H B,  Werner T, Wosnitza J and  L\"ohneysen H v
1996 Phys. Rev. B{\bf 54}  15924
\item{[104]}   Jayasuriya K D,  Campbell S J and  Stewart A M 1985
J. Phys.  F{\bf 15}  225
\item{[105]}   Jayasuriya K D,  Campbell S J and  Stewart A M 1985
Phys. Rev.  B{\bf 31}  6032
\item{[106]}  Jayasuriya K D,  Stewart A M,  Campbell S J and 
Gopal E S R 1984 
J. Phys.  F{\bf 14}  1725
\item{[107]}  See, {\it e.g.\/},  Balberg I and  Maman A 1979
Physica  B{\bf 96}  54
\item{[108]}   Gaulin B D,  Hagen M and  Child H R 1988
J. Phys. (Paris) Colloq. {\bf 49}  327
\item{[109]}   Thurston T R,  Helgesen G,  Gibbs D,  Hill J P, 
B.D. Gaulin and G. Shirane, 1993 
Phys. Rev. B{\bf 70}  3151
\item{[110]}   Gehring P M,  Hirota K,  Majkrzak C F and  Shirane G
1993  Phys. Rev. Lett. {\bf 71}  1087;
Hirota K,  Shirane G,  Gehring P M and  Majkrzak C  1994
Phys. Rev. B{\bf 49}  11967
\item{[111]}   Latarelli M, N\'unez-Rugueiro M D and  Papoular M 1995
Phys. Rev. Lett. {\bf 74}  3840
\item{[112]}   Weinrib A and  Halperin B I 1983 Phys. Rev.
B{\bf 27}  413
\item{[113]}   White G K 1989 J. Phys. C{\bf 1}  6987
\item{[114]}   Tindall D A and  Steinitz M O 1983 
J. Phys. F{\bf 13}  L71
\item{[115]}   Stout J W and  Boo W O J 1966 J. Appl. Phys. {\bf 37}
966;  Stout J W and  Lau H Y 1967
J. Appl. Phys. {\bf 38}  1472;
Lau H Y, Stout J W,  Koehler W C and  Child H R 1969
J. Appl. Phys. {\bf 40}  1136
\item{[116]}   Plumer M L,  Kawamura H and  Caill\'e A 1991
Phys. Rev. B{\bf 43}  13786
\item{[117]}   Visser D,  Coldwell T R,  McIntyre G J,  Graf H,  
Weiss L,  Zeiske Th and  Plumer M L 1994
Ferroelectrics {\bf 162}  147
\item{[118]}   Maleyev S V 1995
Phys. Rev. Lett. {\bf 75}  4682
\item{[119]}   Fedorov V I,  Gukasov A G,  Kozlov V,  Maleyev S V,
Plakhty V P and  Zobkalo I A 1997
Phys. Lett. A{\bf 224}  372
\item{[120]}   Fisher M E and  Nelson D R 1974 
Phys. Rev. Lett. {\bf 32}  1350
\item{[121]}   Kosterlitz J M,  Nelson D R and  Fisher M E 1976 
Phys. Rev. B{\bf 13}  412
\item{[122]}  See, for example,  King A R and  Rohrer H 1979
Phys. Rev. B{\bf 19}  5864
\item{[123]}   Plumer M L,  Hood K and  Caill\'e A 1988 
Phys. Rev. Lett. {\bf 60}  45
\item{[124]}   Plumer M L and  Caill\'e A 1990 
Phys. Rev.  B{\bf 41}  2543
\item{[125]}   Mailhot A,  Plumer M L and  Caill\'e A 1993 
Phys. Rev.  B{\bf 48}  15835
\item{[126]}   Mason T E,  Collins M F and  Gaulin B D 1990
J. Appl. Phys.   {\bf 67}  5421
\item{[127]}   Plumer M L and  Caill\'e A 1990 
Phys. Rev.  B{\bf 42}  10388
\item{[128]}  Poirier M,  Caill\'e, A and  Plumer M L 1990
Phys. Rev. B{\bf 41}  4869
\item{[129]}  Katori H A,  Goto T and  Ajiro Y 1993
J. Phys. Soc. Jpn. {\bf 62}  743
\item{[130]}   Asano T,  Ajiro Y,  Mekata M,  Aruga Katori H 
and  Goto T 1994
Physica B{\bf 201}  75
\item{[131]}   Clark R H and  Moulton W G 1972 
Phys. Rev. B{\bf 5}  788
\item{[132]}   Kadowaki H,  Ubukoshi K and  Hirakawa K 1987
J. Phys. Soc. Jpn. {\bf 56}  751
\item{[133]}   Ajiro Y,  Inami T and  Kadowaki H 1990 
J. Phys. Soc. Jpn.
{\bf 59}  4142
\item{[134]}   Kadowaki H,  Inami T,  Ajiro Y,  Nakajima K
and  Endoh Y 1991
J. Phys. Soc. Jpn. {\bf 56}  1708
\item{[135]}   Beckmann D,  Wosnitza J,  L\"ohneysen H v and  
Visser D 1993
J. Phys. Condens. Matter {\bf 5}  6289
\item{[136]}  Goto T,  Inami T and  Ajiro Y 1990
J. Phys. Soc. Jpn. {\bf 59}  2328
\item{[137]}   Tanaka H,  Nakano H and  Matsuo S 1994
J. Phys. Soc. Jpn. {\bf 63}  3169
\item{[138]}   Weber H,  Beckmann D,  Wosnitza J and  L\"ohneysen H v
1995
Int. J. Mod. Phys.  B{\bf 9}  1387
\item{[139]}  Kawano S,  Ajiro Y and Inami T   1992
J. Magn. Magn. Mater. {\bf 104-107}  791
\item{[140]}  Kato T, Ishii T, Ajiro Y, Asano T and Kawano S 1993
J. Phys. Soc. Jpn. {\bf 62}  3384
\item{[141]}   Heller L, Collins M F, Yang Y S and 
Collier B 1994
Phys. Rev. B{\bf 49}  1104
\item{[142]}   MasonT E,  Stager C V,  Gaulin B D and 
Collins M F 1990
Phys. Rev. B{\bf 42}  2715

\vfil\eject
\noindent
{\bf Figure captions}
\medskip\noindent
\noindent
Fig.1 Ground-state spin configuration of three Ising spins on
a triangle coupled antiferromagnetically. Frustration leads to
the nontrivial degeneracy of the ground state.
\medskip

\noindent
Fig.2 Ground-state spin configuration of three vector spins on a
triangle coupled antiferromagnetically. Frustration leads to the
noncollinear or
canted ordered state. In the case of $n=2$-component {\it XY\/}
spins, the ground state is twofold degenerate according as the
the noncollinear spin structure is either right- or left-handed,
each of which is characterized by the opposite chirality.
\medskip

\noindent
Fig.3 Chiral degeneracy in the ordered state of the {\it XY\/}
antiferromagnet on the triangular lattice.
\medskip

\noindent
Fig.4 Chiral degeneracy in the ordered state of the {\it XY\/}
helimagnet.
\medskip

\noindent
Fig.5 Representations of ``instability points'', 
solid and open circles,
in wavevector space for (a) ferromagnets, (b) antiferromagnets on
bipartite lattices, (c) stacked-triangular antiferromagnets,
and (d) helimagnets. The dashed lines outline the first Brillouin 
zone. Double lines represent the reciprocal lattice vectors $\vec K$:
As usual, points connected by  $\vec K$ should be fully identified.
\medskip

\noindent
Fig.6 Mean-field phase diagram in the $(u,v)$ plane
of the LGW Hamiltonian (2.4), where
$u$ and $v$ are two quartic coupling constants.
On the line $v=4u$, the transition to the noncollinear state is
of mean-field tricritical.
\medskip

\noindent
Fig.7 Renormalization-group flows in the $(u,v)$ plane obtained by
the $\epsilon =4-d$ expansion for the LGW Hamiltonian (2.4).
Parts (a)-(d) correspond to the regimes I-IV specified in the text. 
The hatched regions represent basins of attraction of the stable
fixed point. In (a), the line connecting the Gaussian fixed point $G$
and the unstable antichiral fixed point $C_-$ is the tricritical line
corresponding to the separatrix between 
the two regions in the parameter space,
one associated with a continuous transition (hatched region)
and the other with
a first-order transition. 
\medskip

\noindent
Fig.8 Stability regions in the $(n,d)$ plane, with $\epsilon =4-d$,
of fixed points accessible in the noncollinear region $v>0$.
\medskip

\noindent
Fig.9 Renormalization-group flows in the $(u,v)$ plane in
the noncollinear region $v>0$, 
expected when $n$ is only slightly smaller than $n_{\rm I}(d)$. 
There remains
a ``shadow'' of the slightly complex-valued chiral fixed point which
attracts the flows up to a certain scale.
Eventually, all flows show runaway, signaling a weak
first-order transition.
\medskip

\noindent
Fig.10 Renormalization-group flows in the $(u,v)$ plane in
the noncollinear region $v>0$ just 
at $n=n_{\rm I}(d)$. The hatched regions 
represent basins of attraction of the stable
fixed point. 
The fixed point
$C$ is doubly degenerate, $C_+$ and $C_-$. It is a  
nontrivial fixed point with
a finite domain of attraction in the $(u,v)$ plane.
\medskip

\noindent
Fig.11 Illustration of noncoplanar spin orderings like the
ones realized in triple-$\vec Q$ structures in type-I (above) and in
type-II(below) fcc antiferromagnets. 
\medskip

\noindent
Fig.12 Temperature and size dependence of the
specific heat calculated by Monte Carlo simulation of the
stacked-triangular (a) {\it XY\/} and (b) Heisenberg  
antiferromagnets with
$L^3$ spins. The data are taken from Ref.21. The insets exhibit 
the size dependence of the specific-heat peak.
\medskip

\noindent
Fig.13  Specific heat versus 
reduced temperature $\mid t\mid $
of the
stacked-triangular {\it XY\/} antiferromagnet CsMnBr$_3$,
taken from Ref.27. The inset shows the specific heat in a linear
representation.
\medskip

\noindent
Fig.14   Magnetic Bragg intensity of the 
$(1/3, 1/3, 1)$ reflection 
measured by neutron diffraction for the 
stacked-triangular {\it XY\/} antiferromagnet CsMnBr$_3$
plotted versus reduced temperature.
The data are taken from Ref.25.
\medskip

\noindent
Fig.15 Schematic magnetic field (H) versus temperature (T) phase
diagram of weakly anisotropic unfrustrated antiferromagnet on a
bipartite lattice; (a) axial magnet in a field applied along an
easy axis; (b) planar magnet in a field applied in an easy plane. 
\medskip

\noindent
Fig.16 Schematic magnetic field (H) versus temperature (T) phase
diagram of weakly anisotropic frustrated antiferromagnet on a
stacked-triangular lattice; (a) 
axial magnet in a field applied along an
easy axis; (b) planar magnet in a field applied in an easy plane. 
\medskip

\noindent
Fig.17 Magnetic phase diagram of the axial stacked-triangular
antiferromagnet CsNiCl$_3$ near the multicritical point
as determined by sound-velocity measurements. Labeled regions 1-4
refer to the four phases in Fig.16b.
The data are taken from Ref.128.
\medskip

\vfil\eject
\noindent
{\bf Table captions}
\medskip\noindent
\noindent
Table 1 Order-parameter spaces and the associated homotopy groups for
various continuous spin system in two dimensions.
\medskip

\noindent
Table 2 Critical exponents, amplitude ratio and transition
temperature as determined by several Monte Carlo simulations on 
the stacked-triangular {\it XY\/} antiferromagnet with $J=J'$. 
Maximum lattice size used in each simulation is also shown.
Corresponding values given by several theories are
also shown.
\medskip

\noindent
Table 3 Critical exponents, amplitude ratio and transition
temperature as determined by several Monte Carlo simulations on 
the stacked-triangular Heisenberg antiferromagnet with $J=J'$. 
Maximum lattice size used in each simulation is also shown.
Corresponding values given by several theories are
also shown.
\medskip

\noindent
Table 4 Critical exponents and amplitude ratio  determined by
experiments on several stacked-triangular {\it XY\/} 
antiferromagnets.
The values given by  several theories  are
also shown. 
\medskip

\noindent
Table 5 Critical exponents and amplitude ratio  determined
by experiments on several stacked-triangular Heisenberg
(or nearly Heisenberg) antiferromagnets.
The values given by  several theories  are
also shown. 
Note that VCl$_2$, VBr$_2$ and RbNiCl$_3$ possess a weak
Ising-like anisotropy, which leads to a small splitting of
the transition temperature 
(35.80K and 35.88K in case of VCl$_2$; 11.11K and 11.25K in case of 
RbNiCl$_3$).
Since the fully isotropic critical behavior
should be  interrupted due to the anisotropy
sufficiently close to $T_N$, one should note that
the reported exponents may be affected somewhat
by the crossover effect.

\end